\RequirePackage{fix-cm}
\documentclass[smallextended]{svjour3}           
\smartqed  
\usepackage{graphicx}
\usepackage[round]{natbib}
\setcitestyle{notesep={; },round,aysep={},yysep={;}}
\usepackage{tabularx}
\usepackage{multirow}
\usepackage{bigstrut}
\usepackage{xcolor}
\begin{document}

\title{Cell-based model of the generation and maintenance of the shape and structure of the multi-layered  shoot apical meristem of \textit{Arabidopsis thaliana}
}

\titlerunning{Modeling study of the generation and maintenance...}        

\author{Mikahl Banwarth-Kuhn\textsuperscript{1}\thanks{\noindent\textsuperscript{1}Department of Mathematics, University of California, Riverside}\textsuperscript{,2}\thanks{\noindent\textsuperscript{2} Center for Quantitative Modeling in Biology, University of California, Riverside}\and Ali Nematbakhsh\textsuperscript{1}\textsuperscript{,2}\and Kevin W. Rodriguez \textsuperscript{3}\textsuperscript{,4}\and Stephen Snipes\textsuperscript{3}\thanks{\noindent\textsuperscript{3} Department of Botany and Plant Sciences, Center for Plant Cell Biology (CEPCEB), University of California, Riverside}\textsuperscript{,4}\thanks{\noindent\textsuperscript{4} Institute of Integrative Genome Biology, University of California, Riverside}\and Carolyn G. Rasmussen\textsuperscript{3}\and G. Venugopala Reddy \textsuperscript{3}\textsuperscript{,4}\and Mark Alber\textsuperscript{1}\textsuperscript{,2}}

\authorrunning{M. Banwarth-Kuhn et al.} 

\institute{ Mark Alber \at
         	malber@ucr.edu
 }

\date{Received: date / Accepted: date}

\maketitle

\begin{abstract}
One of the central problems in animal and plant developmental biology is deciphering how chemical and mechanical signals interact within a tissue to produce organs of defined size, shape and function. Cell walls in plants impose a unique constraint on cell expansion since cells are under turgor pressure and do not move relative to one another. Cell wall extensibility and constantly changing distribution of stress on the wall are mechanical properties that vary between individual cells and contribute to rates of expansion and orientation of cell division.  How exactly cell wall mechanical properties influence cell behavior is still largely unknown. To address this problem, a novel, sub-cellular element computational model of growth of stem cells within the multi-layered shoot apical meristem (SAM) of \textit{Arabidopsis thaliana} is developed and calibrated using experimental data.  Novel features of the model include separate, detailed descriptions of cell wall extensibility and mechanical stiffness, deformation of the middle lamella and increase in cytoplasmic pressure generating internal turgor pressure. The model is used to test novel hypothesized mechanisms of formation of the shape and structure of the growing, multilayered SAM based on WUS concentration of individual cells controlling cell growth rates and layer dependent anisotropic mechanical properties of sub-cellular components of individual cells determining anisotropic cell expansion directions.  Model simulations also provide a detailed prediction of distribution of stresses in the growing tissue which can be tested in future experiments.

\end{abstract}
\section{Introduction}
\label{intro}

\begin{figure}
\centering
\includegraphics[width=\textwidth]{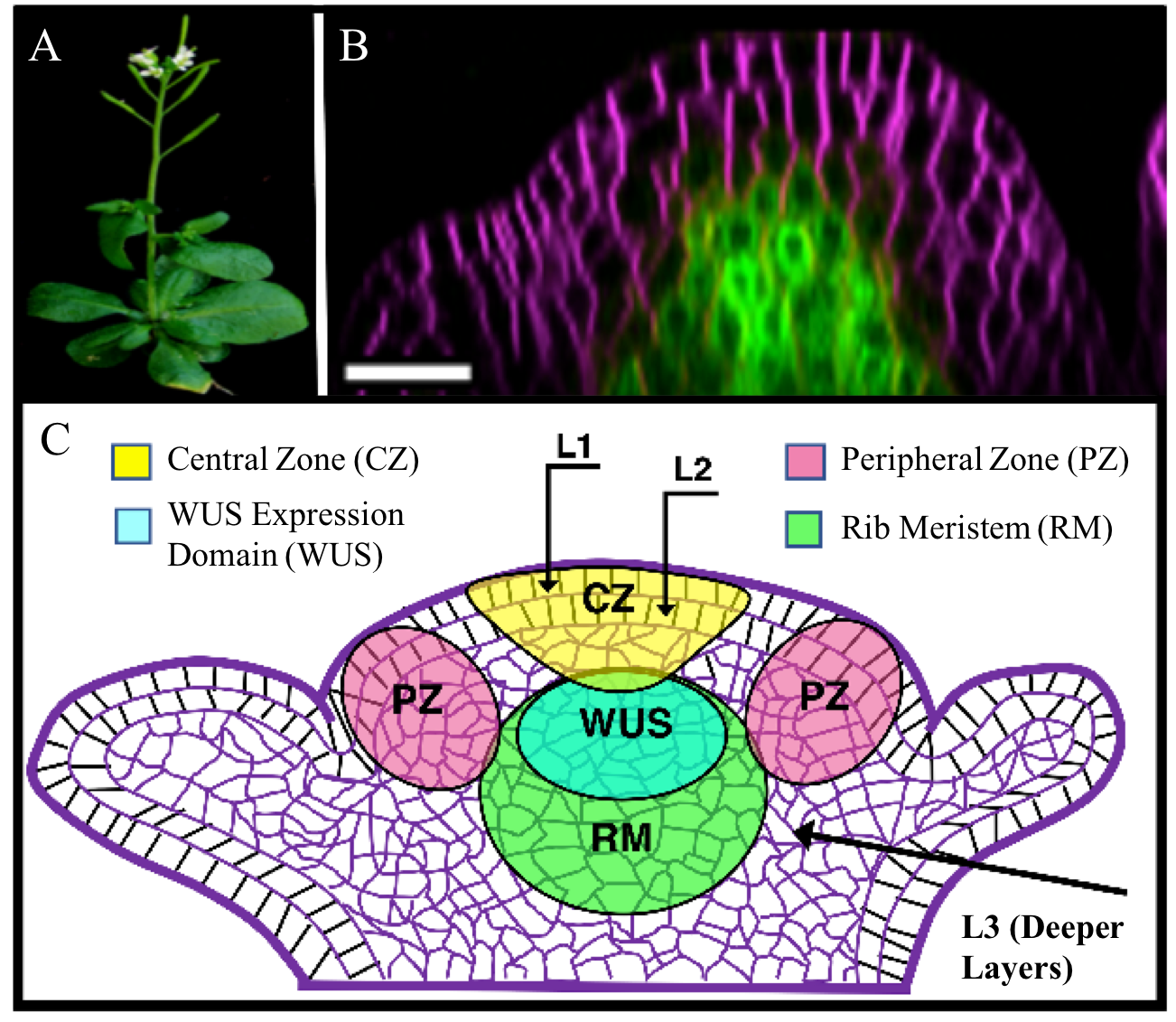}
\caption{Structure and organization of the SAM. (A) The SAM is located at the growing tip of the plant. (B) Higher magnification side experimental image of the SAM showing cell layers, cell boundaries (magenta), and \textit{WUSCHEL (WUS)} expression domain (green) in deeper layers. (C) Diagram showing different functional zones and the three distinct cell layers- L1, L2 and the deeper L3 layers.  Scale bar is 25$\mu$m.}
\label{SAM_Overview}
\end{figure}

One of the biggest questions that faces developmental biology is how molecular signals and physical forces within a developing tissue contribute to its overall form, size, structure and function during morphogenesis.  Unlike their animal counterparts, plant cells do not move relative to one another during development, and the final shape and size of plant tissues or organs are due to coordinated patterns of cell growth, cell wall elongation, cell division, as well as individual cellular response to mechanical stress.  The shoot apical meristems (SAMs) of plants provide an ideal system for studying cell behavior in a morphogenetic and physiological context.  Their essential function is to produce a constant population of stem cells that differentiate into cells for the development of all above-ground organs such as leaves, stems and branches (Figure \ref{SAM_Overview}).

	The SAM in model plant \textit{Arabidopsis} is a multi-layered dome like structure consisting of about 500 cells that is subdivided into different layers and zones (Figure \ref{SAM_Overview} B and C). The outermost L1 layer and the subepidermal L2 layer are single cell layers in which cells divide perpendicular to the SAM surface (anticlinal).  Below the L1 and L2 layers, cells divide both perpendicular to the SAM surface and parallel to the SAM surface (periclinal) to form multiple internal layers collectively called the deeper L3 layers or corpus.  

	Superimposed on this layered organization, cells are also organized by functional zones.  The central zone (CZ) harbors a set of stem cells that span all three cell layers (Figure \ref{SAM_Overview} C). Stem cell progeny are pushed away laterally into the peripheral zone (PZ) where cells divide at a faster rate and differentiate at specific locations to form leaves or flowers.  In addition, stem cell progeny located beneath the CZ in a region termed the rib meristem (RM), also gradually differentiate along the apical-basal axis to form the stem of the plant.  Despite this process of constant displacement and subsequent differentiation, the relative ratios of cells in the CZ, the PZ, and the RM are maintained \citep{Truskina2018}.  This requires a balance between two competing processes, stem cell maintenance and stem cell differentiation.  Each one of these processes is regulated by a set of mechanisms controlling individual cell behaviors such as rate of growth and division, growth direction, and division plane orientation \citep{Lyndon1998,Steeves1989,Xie2009}.  
	
	Molecular and genetic analysis has revealed critical regulators of SAM growth, stem cell maintenance, and organ differentiation \citep{Barton2010,Reddy2004,ReddyandMeyer2005,Yadav2010,Reinhardt2003,Jonsson2006,Smith2006,deReuille2006}.  However, despite the importance of each of these factors in regulating growth and gene expression, our understanding of their feedback mechanisms is incomplete because the underlying dynamics are not well understood.  Early studies show that WUSCHEL (WUS), a homeodomain transcription factor (TF) which is expressed in the RM (Figure \ref{SAM_Overview}C), is responsible for providing cues for stem cell specification in the overlying CZ \citep{Laux1996,Mayer1998}.  

\begin{figure}
\centering
\includegraphics[width=\textwidth]{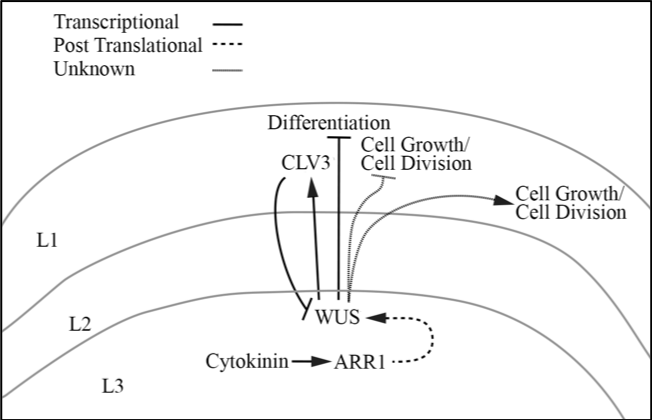}
\caption{Regulatory factors in SAM growth and stem cell maintenance. Cytokinin (CK) signaling stabilizes the WUSCHEL (WUS) protein in the apical L3 layers in the RM likely through activation of TypeB ARABIDOPSIS RESPONSE REGULATOR1 (ARR1).  WUS protein migrates into the CZ where it activates \textit{CLV3} and also represses differentiation-promoting factors.  In the CZ, high levels of WUS decrease cell growth and division rates either directly through an unknown mechanism or indirectly by regulating CZ identity.  Similarly in the PZ, low levels of WUS are associated with an increase in cell growth and division rates.}
\label{Signal}
\end{figure}

	WUS protein migrates from the RM into the overlying CZ and specifies stem cells by repressing differentiation promoting genes (Figure \ref{Signal}) \citep{Yadav2011,Yadav2014}.  In addition, WUS restricts its own transcription by directly activating a negative regulator called \textit{CLV3} (Figure \ref{Signal}) \citep{Fletcher1999,Brand2000,Perales2016}.  \textit{CLV3} encodes a small secreted peptide that activates membrane bound receptor kinases in order to restrict \textit{WUS} transcription in the L1 and L2 layer and reduce \textit{WUS} expression levels in the deeper L3 layers \citep{Clark1997,Ogawa2008}.  Transient depletion of \textit{CLV3} results in radial expansion of the \textit{WUS} expression domain as well as a radial increase in cell division rates among stem cell daughters in the PZ \citep{ReddyandMeyer2005}.   
	
	Additional experiments have shown further that WUS can perform multiple functions depending upon its levels and location of expression.  Misexpression of WUS in the CZ not only induces expansion of the CZ, but also results in increased cell division rates in cells of the PZ where there is low WUS accumulation \citep{Yadav2010}.  Alternatively, over-activation of \textit{CLV3} leads to a smaller CZ and an associated reduction in cell division rates. Classically, this could be correlated to a decrease in WUS levels due to down-regulation of \textit{WUS} transcription \citep{Brand2000, Muller2006}.  
	
	However, recent studies show that  despite higher synthesis of the WUS protein in the RM of \textit{clv3-2} null mutants, these meristems fail to accumulate higher levels of WUS in the CZ \citep{Perales2016}.  This suggests a second function for \textit{CLV3}-mediated signaling in regulating WUS protein levels post-translationally \citeyearpar[For details see Figure 4L in Perales et al.][]{Perales2016}.  The presence of extremely high WUS in the inner layers and extremely low WUS in the outer layers may lead to overproliferation of epidermal cells in the outer layers along with growth restriction of centrally located cells in the deeper layers causing tissue folding and irregular SAMs seen in experiments.  Together, these experiments suggest a more complex regulation of the WUS protein gradient, and indicate that there is no strict correlation between \textit{WUS} transcription and WUS protein accumulation.  
	
	In addition, misexpression of WUS in the CZ results in protein instability that leads to very low, uniform accumulation of WUS and highly enlarged, dome-shaped SAMs \citeyearpar[For details see Figure 5E in Perales et al.][]{Perales2016}.  This suggests that lower WUS accumulation could be responsible for increased growth rates in the PZ as documented in Yadav et al. \citeyearpar{Yadav2010}.  However, an increased number of slow growing cells in the central region of the SAM could either be due to expansion of the CZ identity, or to a transient, higher accumulation of WUS which was not detected in experiments because observations were made at steady-state conditions in terminal SAMs.
	
	Recent experiments reveal that precise accumulation of WUS in space involves several interconnected, intracellular processes such as DNA dependent homodimerization, nuclear retention, and nuclear export which determine nuclear levels that impact WUS protein stability \citep{Rodriguez2016}.  These experiments suggest that the spatial distribution of WUS impacts overall shape and size of the SAM and plays a crucial role in maintaining a constant number of stem cells.  However, the exact impact of WUS levels on cell growth and division patterns in distinct functional domains and how local events influence morphogenetic processes contributing to global tissue patterns that regulate stem cell homeostasis is not well understood. This is because WUS-mediated cell fate specification, and cell growth and division patterns are spatiotemporally coupled.   

	Thus, understanding SAM growth and how it relates to the regulation of stem cell homeostasis, requires the study of dynamic WUS protein regulation leading to its steady-state accumulation, combined with a detailed study of how cells interpret WUS levels to specify cell identity and regulate growth patterns.  The morphological features of the SAM that arise from individual cell behaviors have important physiological implications.  For example, curvature of the L1 layer of the SAM plays a role in determining the distance of the PZ from the RM, and consequently, influences how WUS accumulates in the PZ.  Low WUS accumulation is necessary to allow differentiation and induction of cell division which precede primordium development.  In addition, the shape of the SAM determines distribution of mechanical stress throughout the tissue which plays a role in the establishment of the main axis of expansion of individual cells and subsequently the determination of cell division plane orientation (See section \ref{division} for details). 

	 In this paper, we investigate the morphological implications of individual cell behaviors in the SAM by analyzing the combined impact of WUS concentration of individual cells and mechanical properties of sub-cellular components of individual cells and the cell wall on the shape of the SAM characterized by curvature of the L1 layer.   To do this, we use a novel, cell-based, sub-cellular element (SCE) model.  The general sub-cellular element (SCE) modeling approach has been used before in different biological contexts (See section \ref{general} for details).  The main novelty of this paper is the  extension of the general SCE modeling approach to develop a novel model that enables systematic testing of new hypotheses about the underlying mechanisms driving SAM morphogenesis, as well as application of this model for making specific, biologically-relevant predictions. 

	The novel model combines detailed representations of the following properties of different sub-cellular components as well as  dynamic interactions between them: 1) cell wall mechanical properties controlling anisotropic cell expansion, 2) deformation of the middle lamella of the cell wall, and 3) dynamics of the cytoplasmic pressure to generate turgor pressure.  One of the advantages of the newly developed SCE model is that cellular and cell wall mechanical properties are calibrated directly using experimental data (See section \ref{modelCalib} for details).  

	Model predictive simulations are used to test the novel hypothesized mechanism of formation of the shape and structure of the growing multilayered SAM based on WUS concentration of individual cells controlling cell growth rates and layer dependent anisotropic mechanical properties of sub-cellular components of individual cells determining anisotropic cell expansion directions across the L1, L2 and deeper L3 SAM layers.  Model simulations also provide a detailed prediction of distribution of stresses in the growing tissue which can be tested in future experiments.
	
	The paper is organized as follows. The Methods section starts with a description of the modeling background including a subsection providing a description of the general SCE modeling approach. Then it describes the newly developed SCE model for the multi-layered SAM and provides details of the calibration of single cell model parameters representing mechanical properties, as well as spatial distribution of WUS obtained in experiments.  The Methods section also includes a subsection describing Experimental and Image Analysis Methods.  Next, the model is calibrated using experimental data and used in the Results section to quantify how anisotropic mechanical properties of sub-cellular components of individual cells and the cell wall combined with changes in the diameter of the CZ determine the expansion direction of cells and the degree of curvature of the L1 layer of the SAM.  The paper ends with the Discussion and Conclusions section, where results are summarized and predictions of the model are put in a more general biological context. This section also describes future extensions of the computational modeling environment for simulating the impact of division plane orientation on tissue shape as well as the interaction between mechanical signals and a dynamic signaling model.	

\section{Methods}
\label{methods}

\subsection{Modeling Background}
\label{modelHist}
\subsubsection{Plant Growth and Development}

Multiple modeling approaches have been used to study various aspects of plant growth and development  \citep[For reviews see][]{Ali2014,Chickarmane2010,Jonsson2012,Prusinkiewicz2012}.  The general concept of single cell growth due to cell wall yielding was first formalized by Lockhart \citeyearpar{Lockhart1965}.  He used rate equations for osmotic uptake of water and the irreversible expansion of the cell wall to model one dimensional (1D) elongation of a single plant cell.  In his model, the cell wall is represented as a viscoplastic material that behaves as a rigid body at low stress and flows as a viscous fluid at high stress.  Experimental validation of Lockhart's theory confirmed that single cell growth and expansion in plant cells can be entirely described in terms of internal turgor pressure and the mechanical properties of the cell wall.  Lockhart's model was later extended by Dumais et al.  \citeyearpar{Dumais2006} to account for anisotropic cell wall properties.  This model was used to describe tip growth in cells such as root hairs and pollen tubes.

	More recently, several groups have developed two dimensional (2D) and three dimensional (3D) computational models for simulating growth and expansion of plant tissues \citep{Corson2009a,Corson2009b,DYSON2010,Fozard2013,Merks2011,Mjolsness2006}.  These models incorporate the basic physical principals of single cell growth as well as the mechanical interactions between cells.  Cell-centered models represent individual cells as mass points connected to each other by 1D mechanical elements, such as springs.  This approach has been used to model meristem growth in 3D \citep{Hamant2010,Jonsson2006}.  However, in simulations individual cells were found to slide alongside each other which is never observed in experiments.  Vertex-based models provide a solution for restricted cell movement at low computational cost \citep{Fozard2013}.  In this class of models, each cell is represented as a polygon with edges shared by neighboring cells.  The edges represent the cell walls and are modeled by mechanical elements such as 1D or 2D springs or rods that connect cell vertices in two or three dimensions.  For example, Dupuy et al. model cell walls as 1D beam elements that can be stretched or bent by external loads \citep{Dupuy2006,Dupuy2008,Dupuy2010} .  The strain rate of a beam is directly proportional to the turgor-induced stresses in the walls.  This model was used to analyze the distribution of stresses and strains during the emergence of a primordium at the SAM.  
	
	Another approach used for modeling tissue growth and expansion in plants is the finite-element method (FEM) \citep{Mitchell1982,Niklas1977}.  Hamant et al. \citeyearpar{Hamant2008} used FEM to model stress-strain patterns in the L1 layer of the SAM and correlate results to the behavior of cortical microtubule arrays in individual cells.  In their model, the dome structure of the SAM was represented by a surface made up of 2D polygonal cells in 3D space.  The cells in the deeper layers of the SAM were abstractly represented as a uniform pressure being applied to the surface from below.  Boudon et al. \citeyearpar{Boudon2015} modeled the SAM as a dome structure made from polyhedrons that represent the rigidly connected 3D cells. Cell walls are represented by the faces of the polyhedrons and are composed of 2D elastic triangular elements.  In their model, growth depends on the local modulation of cell wall mechanical properties and turgor pressure.  Using flower development as a case study, Boudon et al. \citeyearpar{Boudon2015} showed how a limited number of gene activities controlling cell wall mechanical properties can explain the complex shape changes that accompany organ outgrowth.

 \subsubsection{General Cell-Based and SCE Modeling Approaches}
\label{general}
	
Recent technological advances in molecular and live-imaging experiments investigating development and growth of multi-cellular tissues provide very large data sets that can be used for the first time to understand how cell-level processes facilitate large-scale tissue properties.  Computational modeling provides a powerful framework that is complementary to experiments and allows for the integration of biochemical and biophysical data from experiments to propose and test novel hypothesized mechanisms of  morphogenesis.  Thus,  models \citep{Pathmanathan2009,Van2015,Loza2017,Milde2014,Christley2010,Nematbakhsh2017} that incorporate individual  behaviors such as cell-cell interactions, polarity in cell growth direction, cell division, differentiation and biochemical signaling events are necessary to quantify the impact of individual cell processes on overall tissue shape, size and function.  For this reason, a class of cell-based modeling approaches has been developed where cells are modeled as discrete entities \citep[For reviews see][]{Fletcher2017, Tanaka2015, Vermolen2012, Bessonov2017}.  

	Unlike continuous descriptions of tissue dynamics, cell-based models can more easily account for individual cell behavior, heterogeneity in mechanical properties of cells and cell-cell interactions. Also, cell-based models can be easily extended to incorporate new biological details at the sub-cellular and cellular levels. As indicated in Fletcher et al. \citeyearpar{Fletcher2017}, cell-based modeling frameworks currently range from vertex models that approximate the membrane of each cell by a polygon, to immersed boundary and sub-cellular element models that allow for more biologically-relevant, emerging cell shapes.  Cell-based models have been successfully used to capture passive biomechanical properties of cells during tissue development and are being extended to investigate the interplay between chemical and mechanical signals in tissue morphogenesis.

	The sub-cellular element (SCE) modeling approach is an established cell-based framework for modeling mechanical properties of individual cells as well as their components and determining their impact on the emerging properties of growing multi-cellular tissue as well as describing cellular interactions with mediums such as the ECM and fluids  \citep{Sandersius2011,Van2015,Loza2017,Tanaka2015,Milde2014,Sandersius2011emerge,Christley2010, Fletcher2017,Brodland2015,Newman2005,Sandersius2008,Newman2007,Amiri2017,Wu2014,Nematbakhsh2017,Sweet2011}.  The general approach was initially developed by Newman et al. \citeyearpar{Newman2005} for simulating the detailed dynamics of cell shapes as an emergent response to mechanical stimuli.  Recent applications of the SCE modeling approach show that it is flexible enough to model additional diverse biological processes such as intercellular chemical kinetics, intercellular signaling, cell differentiation and motion of cells in fluid.

	In the SCE modeling approach, membrane and cytoplasm of each cell are represented using different sets of elements/nodes and their mechanical properties are described using viscoelastic interactions between elements/nodes resulting in coarse-grained molecular dynamics type representation of the cytoskeletal network. Biomechanical and adhesive properties of cells are modeled through viscoelastic interactions between elements represented by phenomenological potential functions that simulate close-range repulsion (modeling volume exclusion of neighboring segments of cytoskeleton) and medium-range attraction  between elements of the same or different cells (modeling the adhesive forces between segments of cytoskeleton)\citep{Morse1929}.  

	One of the important features of the SCE modeling approach is the ability to adjust parameters of potential functions describing connections between elements to calibrate model representations of  biomechanical properties of a particular type of a cell directly using experimental data.  More specifically, the SCE model can be used to perform \textit{in silico} bulk rheology experiments on a single cell in order to scale the parameters such that the passive biomechanical properties of each cell are independent of the number of elements used to represent each cell \citep{Sandersius2008}.  As a result, SCE simulations can capture the underlying biomechanical properties of the real biological system and remain relevant to the real biological system regardless of the number of elements chosen to represent each cell in the model.  

	As indicated in Fletcher et al. \citeyearpar{Fletcher2017}, computational experiments follow a creep-stress protocol in which a constant extensile force is applied to the end of an SCE cell whose opposite end is fixed. Before the extensile force is released, the strain is measured as the extension of the cell in the direction of the force relative to its initial linear size.  \textit{\textit{In silico}} estimates of the viscoelastic properties of cells modeled using the SCE approach have been shown in many biological applications to agree very well with \textit{in vitro} rheology measurements \citep{Sandersius2008,Wottawah2005}.  This indicates that the simple phenomenological dynamics of the SCE modeling approach are enough to capture low to intermediate responses of cytoskeletal networks over short timescales ($\sim 10$s) \citep{Wottawah2005}.  Over longer timescales ($\sim 100$s), cells respond actively to external stresses by undergoing cytoskeletal remodeling, and this phenomenon can be incorporated into the SCE modeling approach by inserting and removing sub-cellular elements of a cell in regions under high or low stress \citep{Sandersius2011emerge}.  
	
	 The SCE modeling approach has been used previously by our group to model platelets in blood stream, and most recently, for studying swarming of bacteria and epithelial cells in an embryo  \citep{Sweet2011,Wu2014,Amiri2017,Nematbakhsh2017}. In this paper, the general SCE modeling approach is applied to develop a novel model that describes combined growth of the L1, L2 and deeper L3 layers of the SAM (See section \ref{modelDescp} for details).  The ability of the SCE model to represent heterogenous mechanical properties on a sub-cellular scale is different from traditional FEM methods that model the entire meristem as one continuous material.  In the FEM and other continuous models, anisotropic properties are defined by assigning different material constants along independent coordinate directions.  As a result, all cells or mesh elements in a tissue have the same mechanical properties.  In some continuous models, small regions of cells or subsets of mesh components are given heterogenous mechanical properties, but there is no variation on the sub-cellular scale.

  The generalized Morse potential functions implemented in our model are commonly used in physics and chemistry to model inter-molecular interactions \citep{Schiff1968} and in biology to represent volume exclusion of neighboring regions of the cytoskeleton \citep{Sweet2011,Wu2014,Amiri2017,Nematbakhsh2017,Christley2010,Gord2014}.  It is difficult to associate specific potential functions directly with specific cytoskeletal components of cells.  However, computational studies of bulk properties at the tissue level have suggested that the precise functional form of the potential used in the model has a small impact on the overall system dynamics \citep{Sandersius2008,Pathmanathan2009}.  An important feature of the SCE modeling approach is the ability to adjust parameters of potential functions describing connections between elements to calibrate model representations of biomechanical properties of a particular type of a cell directly using experimental data \citep{Sandersius2008}.  We used the novel SCE model in our paper to perform simulations of deformation of a single cell to determine parameter values such that the passive biomechanical properties of each cell would be independent of the number of elements used to represent each cell (See section \ref{modelCalib} for details). 

\subsection{Model Description}
\label{modelDescp}

\begin{figure}
\centering
 \includegraphics[width=\textwidth]{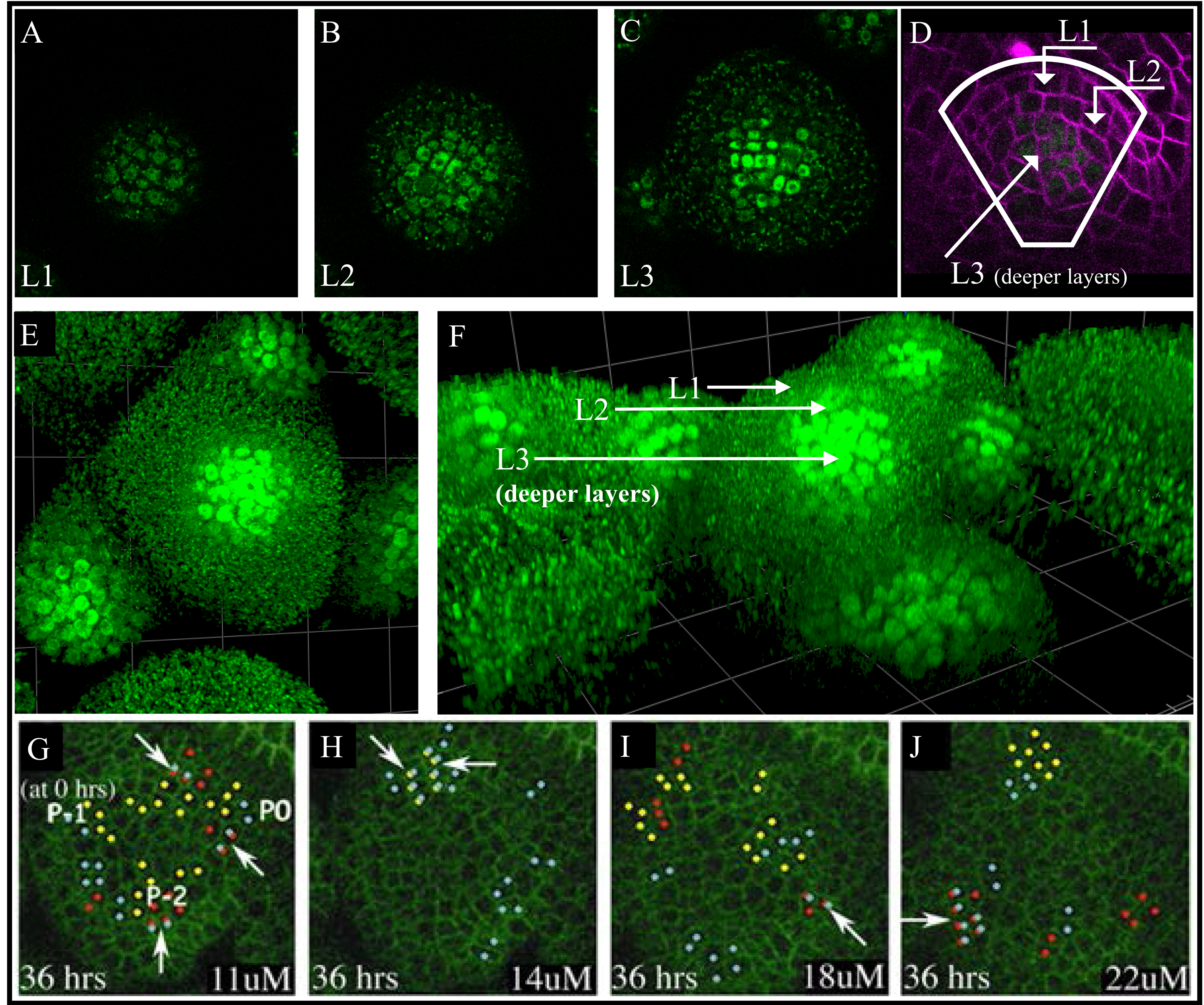}
\caption{Experimental images demonstrating symmetry in distribution of growth rates and the chemical signal WUS in SAM layers as well as dome-like structure of the apical half of the meristem. (A-C) Images show individual top-down sections showing WUS protein accumulation in the L1 (A), L2 (B), and L3 (C) layers (green).  (D) Experimental side-view image showing simulation domain in white. (E-F) 3D reconstructed image of the SAM displaying dome shape of the meristem as well as radial symmetry of WUS signal across L1, L2 and deeper L3 layers (green).  (G-J) Spatial distribution of mitotic activity over time.  Images show individual top-down sections from the same plant, depicting cells located in the L2 and deeper L3 layers at the same time point.  Cells that have divided in each of the 12-hour intervals are color-coded. Red dots represent cells that divided in the first 12-hour window, yellow dots the following 12 hours, and blue dots the final 12 hours.  There is low to no division in the CZ and rates increase as you move toward the PZ.  The overlapping dots indicate a second round of cell division (arrows) which are only present in outermost edge of meristem.  Image reprinted with permission from Reddy et al. \citeyearpar{Reddy2004}.}
\label{symmetry}
\end{figure}

	Our model simulates a 2D longitudinal cross-section of the SAM (Figure \ref{symmetry}).  In Reddy et al. \citeyearpar{Reddy2004} application of the live imaging techniques led to the development of a spatial map of cell growth and division patterns. Cell division rates were found to vary across the SAM surface and it was shown that cell cycle lengths are radially symmetric, i.e. cells in the PZ divide at a faster rate than cells in the CZ (Figure \ref{symmetry} G-J).  In addition, the WUS signaling domain has been shown to be radially symmetric (Figure \ref{symmetry} A-C).  It is important to note that symmetry in growth rates and chemical signaling domains are broken upon formation of organ primordia, but  this happens outside of the domain our model encompasses (Figure \ref{symmetry} D).   
 	
	Experimentally observed symmetry in distribution of growth rates and the WUS signaling domain across SAM layers supports application of a 2D model since it suggests that the apical half of the meristematic dome is radially symmetric, i.e. a longitudinal cut at any angle through the center of the meristem will give the same profile, with respect to both cell growth patterns and the WUS signaling domain within our domain of simulation.  In addition, a 2D model is sufficient to predict shape, quantified by curvature of the L1 layer, since the dome-like structure of the apical half of the meristem ensures curvature of the L1 layer in longitudinal cross-sections of the SAM will be invariant under the choice of angle of the cut within our simulation domain (Figure \ref{symmetry} E-F).
 
	In what follows we first describe different types of sub-cellular nodes that are used to simulate different components of  each cell and the cell wall as well as the potential functions describing interactions between them.  Then we describe the approach implemented for modeling cell growth, cell wall elongation as well as cell division and for determining anisotropic mechanical properties to provide a complete model description.  Finally, the equations of motion for each sub-cellular element are provided along with the numerical method used to solve them. 

\subsubsection{Turgor Pressure}
\label{turgor}
	
\begin{figure}
\centering
 \includegraphics[width=\textwidth]{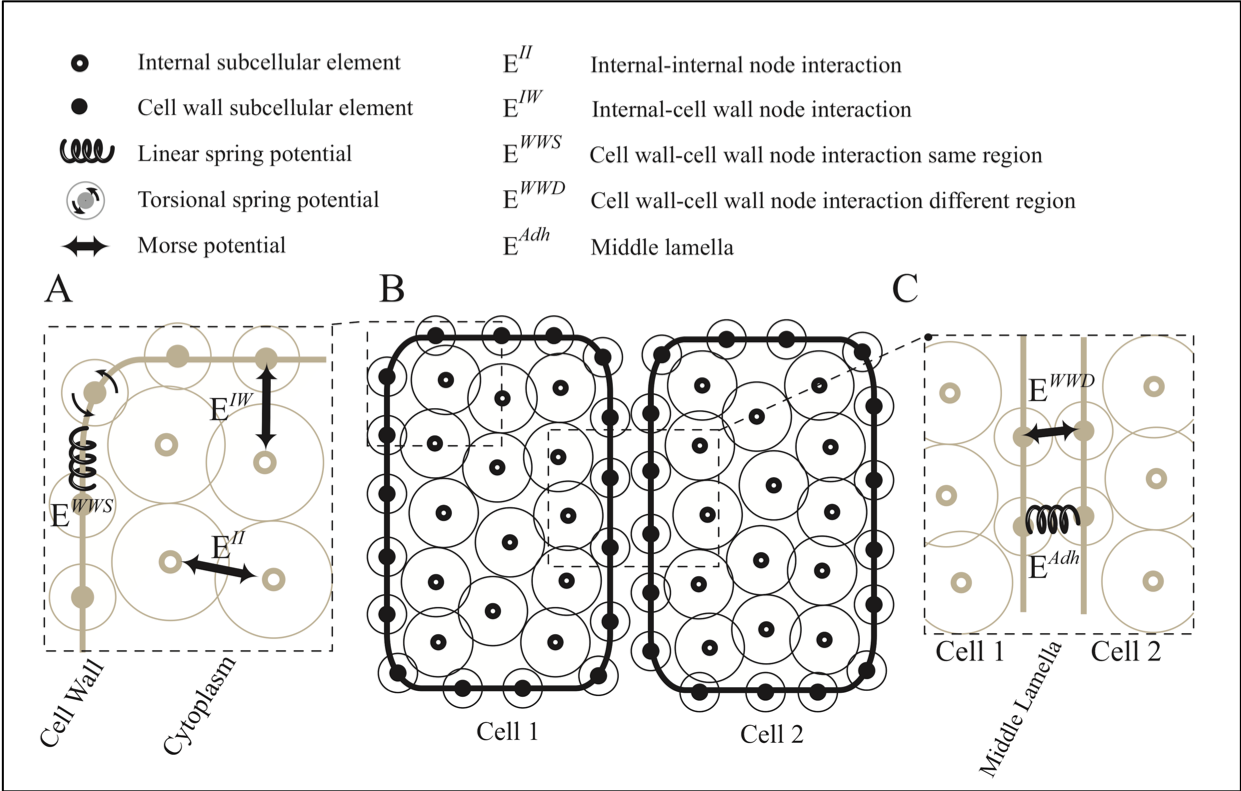}
\caption{Diagram of interactions of the SCE model components represented by different types of nodes.  (A) Intracellular interactions between cytoplasm and primary cell wall nodes.  (B) SCE model components of two neighboring cells. (C) Diagram of the intercellular interactions between two neighboring cells involving middle lamella.  Symbols and notations are described in the figure itself.}
\label{SCE_diagram}
\end{figure}

	Unlike animal cells, each plant cell consists of a "membrane bag" or protoplast sitting inside the cell wall, a mechanically strong and dynamic extracellular matrix that is deposited by the cell outside of its plasma membrane.  In our model, two groups of nodes are used to represent the cell wall and internal cell domains separately (Figure \ref{SCE_diagram}).  Collective interactions between pairs of internal nodes $(E^{II})$ represent the cytoplasmic pressure of the cell, and collective interactions between pairs of internal nodes and primary cell wall nodes $(E^{IW})$ represent turgor pressure, the force per unit surface applied on the cell wall by the protoplast (Figure \ref{SCE_diagram} A). 

\begin{table}
\renewcommand{\arraystretch}{1.25}
\caption{Potential energy functions in the model}
\label{tab:1}       
\begin{tabularx}{\textwidth}{| X | X | X |}
\hline
\textbf{Elements of the SCE model} & \textbf{Type of potential acting on each element} & \textbf{Biological feature}  \\
\hline
Internal-internal node ($E^{II}$) & Morse potential & Internal pressure \\
\hline
Internal-cell wall node ($E^{IW}$) & Morse potential & Turgor pressure, the force per unit surface applied on the primary cell wall by the protoplast \\
\hline
Cell wall-cell wall node in the same cell wall region ($E^{WWS}$) & Linear and rotational spring potential functions & Mechanical stiffness and extensibility of the primary cell wall\\
\hline
Cell wall-cell wall node of neighboring cells ($E^{WWD}$) & Morse potential &  Volume exclusion of the cells due to cell wall material such as cellulose microfibrils and pectin that sit between neighboring cells and keep adjacent cell membranes from making contact \\
\hline
Cell wall-cell wall node of neighboring cells ($E^{Adh}$) & Linear spring potential function & Middle lamella\\
\hline
\end{tabularx}
\end{table}

	Plant cells are under high internal turgor pressure, generally in the range of $0.1-1$ MPa \citep{Geitmann2006}, and are prevented from bursting by the presence of the cell wall.  Turgor pressure is generated when water crosses the cell membrane by osmosis, and causes the protoplast (cell excluding the cell wall) to swell. Swelling of the protoplast is restricted by the cell wall and this generates turgor pressure. These interactions between pairs of internal nodes, and pairs of internal and primary cell wall nodes are modeled using Morse potential functions. 

	The Morse potential used in the model consists of two terms generating short-range repulsive and long-range attractive forces.  The following equation is a Morse potential which models the interaction between internal node $i$ and cell wall node $j$:
\begin{equation}
E_{ij}^{IW} = \left[U^{IW}\exp\left(-\frac{|x_i -x_j|}{\xi^{IW}}\right) - W^{IW}\exp\left(-\frac{|x_i -x_j|}{\gamma^{IW}}\right)\right]
\end{equation}
where $U^{IW}, W^{IW}, \xi^{IW},$ and $\gamma^{IW}$ are Morse parameters.  The same form of the potential with different sets of parameters is used for $E^{II}$ and $E^{WWD}$ (Table \ref{tab:1} and Table \ref{tab:2}).

\subsubsection{Cell Wall and Middle Lamella}
\label{cell wall}

	In plants, the primary cell wall is  composed of cellulose microfibrils cross-linked by a network of polysaccharides, including hemicelluloses and pectins \citep{Cosgrove2001, Daher2015, Smith2001,Liu2015}.  The plasma membrane of individual cells is tightly attached to the adjacent primary cell wall region through transmembrane proteins and sensors on the plasma membrane act as signals for to the cell to export new material and facilitate cell wall remodeling \citep{Liu2015}.  The plasma membrane provides a physical barrier between the cell and the primary cell wall but does not add additional mechanical strength \citep{Liu2015}.  As such, primary cell wall nodes in our model represent  mechanical properties of the primary cell wall and plasma membrane together (Figure \ref{SCE_diagram}).
	
\begin{figure}
\centering
 \includegraphics[width=.55\textwidth]{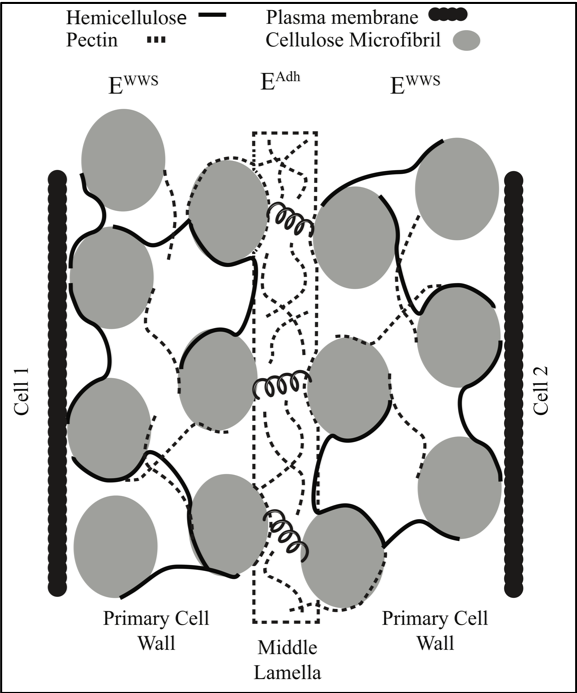}
\caption{Diagram of three cell wall regions between two neighboring cells: two primary cell wall regions on either side of the middle lamella.}
\label{Middle_lam}
\end{figure}

	The cell wall separating two neighboring cells can be viewed as having three separate regions \citep{Daher2015}, two primary cell wall regions immediately adjacent to the plasma membrane of each cell sitting on either side of the middle lamella (Figure \ref{Middle_lam}).  In our model, we represent each region of the primary cell wall with an individual set of nodes surrounding each cell.  Interactions between primary cell wall nodes of the same cell $(E^{WWS})$ are used to model cell wall mechanical stiffness and extensibility (Figure \ref{SCE_diagram} and Table \ref{tab:1}).  There are two types of interactions between primary cell wall nodes of neighboring cells (Figure \ref{SCE_diagram}C and Figure \ref{Middle_lam}).  $(E^{WWD})$ is a repulsive force that is modeled using a Morse potential function to prevent membranes of adjacent cells from overlapping.  This represents cell wall material present between neighboring cells which keeps adjacent cells membranes from making contact.  Pairwise linear spring interactions are used to model cell-cell adhesion mediated through the middle lamella  (Figure \ref{SCE_diagram} C, \ref{Middle_lam} and Table \ref{tab:1}). 

	In addition to molecular signaling discussed in the introduction, plant cells also respond to mechanical forces.  One of the primary forces acting on the plant cell wall is generated by the internal turgor pressure which is strictly isotropic.  However, given that plant cells often expand faster in one direction over the other \citep{Baskin2005},  cell wall resistance to stress could be anisotropic.  This is largely due to the reinforcement of primary cell walls by rigid cellulose microfibrils that have tensile strength comparable to steel \citep{NCBI,Baskin2005}.  Cellulose microfibrils are long, filamentous structures, directly polymerized at the interface of the cell wall and plasma membrane by transmembrane cellulose synthase complexes.  In cells with a preferred growth direction, adjacent cellulose microfibrils are deposited into the wall in such a way that they align parallel to one another and form bundles. 

	The orientation and level of alignment of microfibril bundles within the cell wall is often equated to cell wall resistance since the direction of maximal expansion of the cell is perpendicular to the net orientation direction of the microfibril bundles \citep{Baskin2005}. Cortical microtubules (CMTs) guide the deposition of cellulose microfibrils (CMTs) into the cell wall \citep{Paredez2006}. Recent literature \citep{Hamant2008,Sampathkumar2014a,Sampathkumar2014b, Uyttewaal2012,Williamson1990} provides evidence that microtubules in plant cells align along the main stress direction of the cell and therefore cellulose microfibrils are deposited into the cell wall along this same well-defined direction. As the fibrils are laid down, growth in that direction decreases \citep{Paredez2006}.  In this way, plant cells have the ability to act autonomously to modify and structurally reorganize the primary cell wall to control anisotropic deformation suggesting that mechanical stress feeds back into individual cell behaviors such as anisotropic expansion direction and division plane orientation that control overall shape and size of the tissue.

	Several modeling studies have investigated the importance of microtubule dynamics in cell growth \citep{Burian2013,Allard2010}.  Since CMTs do not contribute directly to cell wall resistance to stress \citep{Lockhart1965}, we developed a coarse-grained model that represents cellulose microfibrils and CMT dynamics through motion of nodes connected by linear and rotational springs.  Namely, interactions that lead to anisotropic expansion through modification and structural reorganization of the primary cell wall are represented by linear and rotational springs ($E^{WWS}$) (Figure \ref{SCE_diagram}A).  Linear-spring interactions given by the following equation $E_{linear} = \frac{1}{2}k_{linear}(x-x_{eq})^2$ are defined between adjacent nodes of the cell wall to maintain the length of cell wall segments and regulate cell wall extensibility (See section \ref{cell_wall_growth}).  Rotational spring interactions defined between three successive nodes of the cell wall are described by the following equation, $E_{bend} = \frac{1}{2}k_{bend}(\theta - \theta_{eq})^2$  and are used to maintain a prescribed degree of bending between cell wall segments \citep{Bathe1984}.  The degree of bending between cell wall segments represents the level of alignment and coordinated orientation of the cellulose microfibrils, and parameters of the bending equation were chosen to mimic cell shape observed in experimental images.  Bending stiffness in the model limits cell expansion along the axis perpendicular to the preferred growth direction, similar to how cells lay down microfibrils to limit expansion in the experimental observations.   The parameters $k_{bend}$ and $k_{linear}$ were calibrated using elastic modulus of cells measured in experiments (See section \ref{modelCalib}).  In simulations, spring constants of primary cell wall nodes are varied based on cell layer and prescribed growth direction of cells leading to anisotropic mechanical properties of the cell wall.
	 	
	In addition to providing mechanical strength, the cell wall also mediates cell-cell adhesion through the pectin-rich middle lamella \citep{Daher2015, Smith2001,Liu2015}.  The middle lamella is primarily composed of pectin, a group of complex polysaccharide-molecules that cross-link the primary cell walls of neighboring cells.  Adjacent pectin chains are cross-linked by calcium ions which facilitates cell-cell adhesion in plants \citep{Daher2015, Smith2001,Liu2015}.  In our model, pairwise interactions between cell wall nodes of adjacent cells $(E^{Adh})$ function as a coarse-grained model for cross-linking of pectin molecules in the middle lamella (Figure \ref{SCE_diagram}C, \ref{Middle_lam} and Table \ref{tab:1}).

\subsubsection{Cell Growth and Anisotropic Cell Wall Expansion}
\label{cell_wall_growth}
	
	Live imaging techniques were previously used to analyze cell cycle lengths in real-time and generate a spatial and temporal map of cell growth and division patterns in the SAMs of \textit{Arabidopsis} \citep{Reddy2004}. In the current model, cell growth is represented by the addition of new internal nodes at a constant rate (Figure \ref{Calibration_image} and section \ref{modelCalib}).  When a new internal node is added, the internal area of the cell increases as nodes readjust to achieve their new equilibrium arrangement.  As the internal area of a cell increases, the cell wall will stretch.  When the cell wall becomes stretched enough that the distance between two successive cell wall nodes passes the membrane threshold length, $\textrm{Linear}_{thresh}$ (See Table \ref{tab:2}), a new cell wall node is added.  This is how cell wall elongation is achieved.  
	
	The addition of new cell wall nodes in the model represents the addition of new cell wall material in the biological system.  When stretched above a certain threshold, the pectin cross-links in the cell wall break and the insertion of new cell wall material results in the irreversible expansion of the cell wall.  The addition of new cell wall material and formation of new pectin cross-links allows cells to increase their size without compromising strength of the cell wall.  Thus, modeling growth as elastic stretching combined with the addition of new cell wall material is a biologically-relevant component of our model since wall expansion due to turgor pressure is accompanied by the synthesis and integration of new wall material \citep{Cosgrove2005,Daher2015, Smith2001,Liu2015}.  

	In addition, since nodes function as a coarse-grained representation of cell wall material, new nodes are added in simulations to maintain the resolution scale (See section \ref{coarse-grained}).  Representation of the two primary cell wall regions on either side of the middle lamella is also a biologically relevant component of our model since contribution of new cell wall material during expansion is carried out independently by neighboring cells, and the orientation and rate of microfibril deposition can vary between adjacent cells \citep{Uyttewaal2012}.  Moreover, individual cell wall mechanical parameters including extensibility and mechanical stiffness play an important role in determining the rate of cell expansion, the main axis of expansion of a cell, and consequently the degree of growth anisotropy.  How individual cells regulate these parameters is fundamental to understanding how plants control global tissue patterns \citep{Coen2004,Erickson1976,Kennaway2011}.
	
	In the SAMs of \textit{Arabidopsis}, the main axis of expansion of cells varies between the different cell layers likely due to differences in anisotropic cell wall properties.  In our model, linear and rotational spring parameters regulate cell expansion directions by controlling the degree of wall extensibility along each axis.  To do this, a growth direction vector, $d_c$, is defined for each cell upon creation and remains unchanged throughout each simulation.  Then, a cell wall direction vector, $d_w$, is computed for each section of the wall.  The cell wall direction vector is defined to be the vector connecting two successive nodes. 

	 Finally, the linear spring constant for each section of the cell wall is determined as a function of $d_c$ and $d_w$ in the following way
\begin{equation}
k_{linear} = k^{min}_{linear} + k^{max}_{linear}(1-\cos^2\theta)
\end{equation}
 where $\theta$ is the angle between the growth direction vector, $d_c$, and the cell wall direction vector, $d_w$, $k_{linear}^{min}$ is the minimum value for the linear spring constant of a cell wall section, and $k_{linear}^{max}$ is the maximum linear spring constant of a cell wall section (See Table \ref{tab:2}).  Cell wall segments assigned lower linear spring constants will stretch apart more easily, facilitating faster expansion in the direction parallel the growth direction vector for that cell.

 \subsubsection{Division}
 \label{division}
Regulation of division plane orientation of individual cells is one mechanism multicellular organisms use to control the shape of their tissues.  Previous experimental studies have revealed a link between tension and division plane orientation in plant cells \citep{Louveaux2016, Louveaux2013}.  Before plant cells enter mitosis, cortical microtubules reorganize into a ring called the preprophase band (PPB), that determines the position of the new cell wall \citep{Rasmussen2013}.  Accumulating experimental evidence suggests that cortical microtubules align along the direction of maximal tensile stress in cell walls, implying that cortical microtubules may play an important role in cell interpretation of tension patterns in cell walls to determine cell division plane orientation \citep{Hamant2008, Uyttewaal2012,Sampathkumar2014}.  Additionally, RM-localized CK may promote periclinal cell divisions.

	Experimentally tracking division plane orientation along with maximal tensile stress in cell walls and level of biochemical signaling is difficult, especially in cells that are located in the deeper layers of the SAM.  For this reason, we model several possible mechanisms driving the positioning of the new cell wall during cell division.  Cell division in simulations occurs once the number of internal nodes has doubled.  Two spots on opposite sides of the cell wall are chosen and the cell division plane is then determined as the plane that goes through the cells' center of mass and connects these two spots on the wall.   The cell is then divided by a straight line created from a set of new cell wall nodes.  After division, parameters for nodes of each individual daughter cell are inherited from the divided cell, and each daughter cell starts with half the amount of cytoplasm that was in the divided cell.  

	Division plane orientation in simulations is determined based on several possible mechanisms.  The cell will determine the position of the new cell wall according to tensile stress in its cell wall, according to its concentration of CK signaling, or a combination of both the mechanical and biochemical signals it's experiencing.  Division plane orientation based on maximum tensile stress in the cell wall only is determined by choosing the two pairs of adjacent cell wall nodes that are furthest apart since these are the spots where the cell wall is under highest tensile stress.  

Alternatively, division plane orientation based on chemical signaling only  is determined by the level of CK concentration in the cell.  Cells with CK concentration above a certain threshold will divide periclinally regardless of mechanical stress on the cell wall. Both mechanisms for division will be tested in future simulations to determine the relative contribution of mechanical stress and CK concentration in determining division plane orientation.  The effect of division plane orientation on morphological features such as cell growth direction and curvature of the L1 layer of the SAM will be compared against experimental images and used to determine the contribution of each type of signal in determining division plane orientation.

\subsubsection{Equations of Motion}
\label{motion}

The potential functions described above are used in the model equations to calculate the displacement of each internal or cell wall node at each time step based on their interactions with neighboring nodes resulting in the deformation of cells within the tissue.  A complete list of all potential functions, parameters, and their biological relevance are provided in Table \ref{tab:1} and Table \ref{tab:2}.  We assume that the nodes are in an overdamped regime so that inertia forces acting on the nodes are neglected \citep{Farhadifar2007,Kursawe2015,Newman2005}. This leads to the following two equations of motion describing the movement of internal nodes and cell wall nodes respectively:
\begin{equation}
\eta\dot{x}^I_{i} = - \left( \sum_j \nabla E_{ij}^{IW}  + \sum_m \nabla E_{im}^{II} \right) 
\end{equation}
\begin{equation}
\eta\dot{x}^W_{j} = - \left( \sum_i \nabla E_{ij}^{IW}  + \sum_k \nabla E_{kj}^{WWS} + \sum_l \nabla E_{lj}^{WWD} + \nabla E_j^{Adh} \right)
\end{equation}
where $i = 1,2,...,N^I$ represent all internal nodes and $j = 1,2,...,N^W$ represent all cell wall nodes.  $\eta$ is the damping coefficient, $x_i^I$ and $x_j^W$ are positions of internal nodes and cell wall nodes indicated by indices $i$ and $j$ respectively, $m$ is the index for any internal node interacting with internal node $i$, $k$ is the index for any cell wall node of the same cell interacting with cell wall node $j$, and $l$ is the index for any cell wall node of a different cell interacting with the cell wall node $j$.  The two equations above are solved at the same time for all internal and cell wall nodes.

The two equations  are discretized in time using the forward Euler method and positions of nodes $x_i^I$ and $x_j^W$ are incremented at discrete times as follows:

\begin{equation}
x_i^{I}(t+\Delta t) = x_i^{I}(t) - \left( \sum_j \nabla E_{ij}^{IW}(t) + \sum_m \nabla E^{II}_{im}(t)\right)\frac{\Delta t}{\eta}
\end{equation}
where $\Delta t$ is the time step size.  The same discretization technique is used for the equations of motion of the cell wall nodes.

\begin{table}[hbtp]
\renewcommand{\arraystretch}{1.25}
\centering
  \caption{Parameter values used in simulations}
    \begin{tabular}{|c|c|c| p{5cm}}
	 \hline
    \multicolumn{1}{|c|}{\textbf{Potential Function}} & {\textbf{Parameter}} & \multicolumn{1}{c|}{\textbf{Value}} \bigstrut\\
    \hline
    \multirow{4}[8]{*}{$E^{II}$} & $U^{II}$   & 75  $nN.\mu m$ \\
\cline{2-3}          & $W^{II}$   &   6.71  $nN.\mu m$ \bigstrut\\
\cline{2-3}          & $\xi^{II}$ & 0.8  $nN.\mu m$ \bigstrut\\
\cline{2-3}          & $\gamma^{II}$ &    1.34  $nN.\mu m$ \bigstrut\\
    \hline
    \multirow{4}[8]{*}{$E^{IW}$} & $U^{IW}$   &  45  $nN.\mu m$  \bigstrut\\
\cline{2-3}          & $W^{IW}$   &   0  $nN.\mu m$   \bigstrut\\
\cline{2-3}          & $\xi^{IW}$ &  0.3  $nN.\mu m$ \bigstrut\\
\cline{2-3}          & $\gamma^{IW}$  &   0 $nN.\mu m$ \bigstrut\\
    \hline
    \multirow{4}[8]{*}{$E^{WWD}$} & $U^{WWD}$   &   3.9  $nN.\mu m$  \bigstrut\\
\cline{2-3}          & $W^{WWD}$   &   0  $nN.\mu m$    \bigstrut\\
\cline{2-3}          & $\xi^{WWD}$ &    0.5  $nN.\mu m$  \bigstrut\\
\cline{2-3}          & $\gamma^{WWD}$ &   0  $nN.\mu m$    \bigstrut\\
    \hline
\multirow{3}[4]{*}{$E^{Adh}$} & $k^{Adh}$     &   20  $nN.\mu m$   \bigstrut\\
\cline{2-3}          & $L^{Adh}$    &  0.8  $\mu m$    \bigstrut\\
\cline{2-3}          & $\textrm{Adh}_{thresh}$    &  2  $\mu m$    \bigstrut\\
   
    \hline
    \multirow{5}[6]{*}{$E^{WWS}$} & $k_{linear}$  &    150-800   $nN.\mu m$  \bigstrut \\
\cline{2-3}         & $k_{min}$ & 150 $nN.\mu m$ \bigstrut\\
\cline{2-3}         & $k_{max}$ & 500 $nN.\mu m$ \bigstrut\\
\cline{2-3}          & $x_{eq}$     & 0.07  $\mu m$      \bigstrut\\
\cline{2-3}          & $k_{bend}$  &    12  $nN.\mu m$ \bigstrut\\
\cline{2-3}          & $\theta_{eq}$  &    circle \bigstrut\\

\cline{2-3}         & $\textrm{Linear}_{thresh}$ & .15 $\mu m$ \bigstrut\\
    \hline
    \multirow{1}[3]{*}{} &  \multicolumn{1}{c|}{$\eta_{stem}$} & \multicolumn{1}{c|}{3} \bigstrut\\
    \hline
     \multirow{1}[3]{*}{} &  \multicolumn{1}{c|}{$\eta_{normal}$} & \multicolumn{1}{c|}{1} \bigstrut\\
    \hline
\multirow{1}[3]{*}{} &  \multicolumn{1}{c|}{Initial number of internal nodes} & \multicolumn{1}{c|}{15} \bigstrut\\
\hline
 \multirow{1}[3]{*}{} &  \multicolumn{1}{c|}{Initial number of cell wall nodes} &  \multicolumn{1}{c|}{150} \bigstrut\\
\hline
  \multirow{1}[3]{*}{} &  \multicolumn{1}{c|}{Time step} & \multicolumn{1}{c|}{0.003}  \bigstrut\\
\hline
    \end{tabular}%
  \label{tab:2}%
\end{table}%

\subsubsection{Model Components at Different Scales}
\label{multiscale}

Our model is multi-scale in space and combines four different scales for modeling growth of the meristem.  Molecular level descriptions include cell-cell adhesion achieved through coarse-grained approximation of pectin cross-linking in the middle lamella and growth rate determined by WUS concentration of each cell (See section \ref{modelCalib}).  Sub-cellular level descriptions include separate node representations of the mechanical properties of individual sub-cellular components of the cell wall resulting in detailed simulation of cell growth and anisotropic cell wall expansion and sub-cellular representation of increase in cytoplasmic pressure to generate turgor pressure resulting in detailed simulation of interaction of cytoplasm and cell wall.  

	Cell level descriptions include detailed description of individual cell behavior including determination of the cell growth direction and interactions of neighboring cells modeled through modification and structural reorganization of the cell wall and cell-cell adhesion.  Descriptions of behavior at the multicellular, tissue level include response of the tissue to non-homogeneous distribution of WUS protein, multicellular interactions between the three different cell layers that lead to shape and size of the meristem (See section \ref{baseline} and \ref{cellGrowth}) and model provides a detailed description of stresses in tissue (See section \ref{stress}).

\subsubsection{Coarse Graining Approach}
\label{coarse-grained}

\begin{figure}
\centering
 \includegraphics[width = \textwidth]{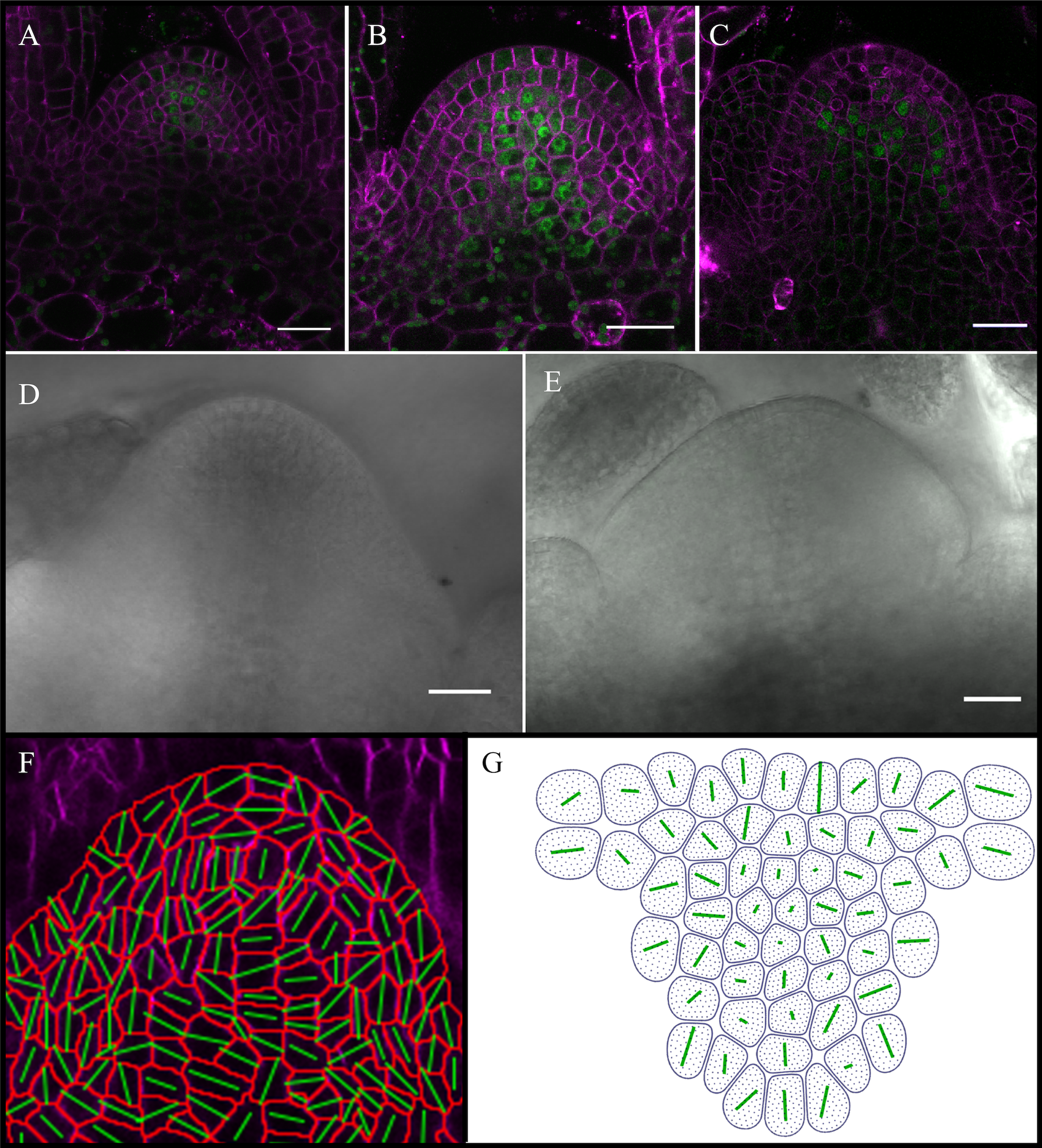}
\caption{Experimental images from wildtype and four alternative systems and application of the image quantification methods. (A) Wildtype SAM showing WUS accumulation in green. Reprinted with permission from \citep{Snipes2018}. (B) Meristem experiencing the ectopic overactivation of CK signaling in the CZ for 12 hours.  Increased WUS accumulation shown in green. Reprinted with permission from \citep{Snipes2018}.(C) \textit{clv3-2} null mutants obtained by our group. (D) Ectopic activation of eGFP-WUS from the CZ-specific \textit{CLV3} promoter.  Reprinted with permission from \citep{Perales2016}.  (E) Misexpressed eGFP-WUS form, in the CZ, that is tagged with a potent nuclear localization signal (nls-eGFP-WUS). Reprinted with permission from \citep{Perales2016}.  (F-G) Main axis of expansion of cells in wildtype SAM from experiments (F) and simulations (G). Green bars depict the main axis of expansion calculated for each cell. Scale bars are $20\mu$m.}
\label{Exp_images}      
\end{figure}

 In our simulations, the number of nodes used to represent each cell is chosen based on the desired level of coarse graining representation.  Then, Morse potential parameters are calibrated based on the average size of cells determined from experimental images (Figure \ref{Exp_images}). Next, the number of cell wall nodes is chosen to make sure volume exclusion is satisfied.  Finally, we wanted the minimum number of elements that met these criteria for computational considerations.  Cell wall nodes in the beginning of a simulation are arranged in a circle for each cell, and internal nodes are randomly placed within each cell. After initialization, internal nodes rearrange and cells attain biological shapes, similar to the experimentally observed cell shapes in the SAM (Figure \ref{Time_series}).  Cells in a simulation constantly grow and interact with each other resulting in a detailed dynamic representation of the combined growth of the L1, L2 and deeper L3 layers of the SAM tissue.
 
\subsection{Experimental and Image Analysis Methods}
\label{expMethods}

\subsubsection{Experimental Methods}
Side view experimental images of sectioned SAMs were obtained by confocal microscopy \citep{Snipes2018}.  SAMs were imaged using a fusion protein of eGFP-WUS to track WUS accumulation \citep{Snipes2018,Yadav2011}.  In addition, plasma membrane staining was used to provide a proxy for visualization of individual cell walls \citep{Snipes2018}.  Wildtype plants used for model validation were grown under normal conditions. 

	To study spatial manipulation of WUS levels, four systems were employed: 
	
	1) A recent study has shown that CK signaling stabilizes the WUS protein in the deeper L3 layers of the RM \citep{Snipes2018}. To induce cytokinin response in cells of the L1 and the L2 layers, active TypeB ARABIDOPSIS RESPONSE REGULATOR1 (ARR1), a transcription factor that functions downstream of the CK receptors, was constitutively misexpressed in dexamethasone inducible fashion, by using the \textit{CLV3} promoter \citep[For further details see][]{Snipes2018}.  For this experiment, ectopic activation of CK signaling in the outer layers leads to an increase in the diameter of the WUS signaling domain as well as increased WUS accumulation in the meristem that spreads out into the deeper layers, and modestly into the L1 or L2 layers (Figure \ref{Exp_images}B) \citep{Snipes2018}. 
	
	2) Ectopic activation of eGFP-WUS from the CZ-specific \textit{CLV3} promoter leads to uniformly lower WUS accumulation in all cell layers of highly enlarged and much taller SAMs \citep{Yadav2010,Perales2016} (Figure \ref{Exp_images} D). 
	
	 3) To achieve higher levels of nuclear WUS, we utilized data sets from an earlier study which misexpressed an eGFP-WUS form, in the CZ, that is tagged with a potent nuclear localization signal (nls-eGFP-WUS) (Figure \ref{Exp_images} E).  For further details see figure 5C and F in Perales et al. 2016.  In this condition, higher nuclear WUS was detected in patches of cells in highly irregularly shaped and much flatter SAMs. 
	 
	 4) WUS accumulation was followed in \textit{clv3-2} null mutants which accumulate WUS at much higher levels in the nuclei of L2 and deeper L3 layers and extremely low levels in the nuclei of the L1 layer (Figure \ref{Exp_images} C) \citep{Perales2016}. 

\subsubsection{Image analysis}

Analysis and quantification of the WUS signal was performed using a combination of ImageJ and the HK-means and Active contour packages within the ICY bio-image analysis software \citep{Snipes2018}.  Plasma membrane staining makes it possible to distinguish between the cell outlines of individual cells and thus measure the amount of signal in each cell as well as describe other cell characteristics such as cell center and the main axis of expansion for each cell.

	The main axis of expansion of cells in both experimental images and simulations is quantified for comparison and model validation.  For \textit{in vivo} cells, the main axis of expansion is inferred from cell shapes observed from single-time-point images.  First, images are segmented in ImageJ.  Next, the EpiTools image processing software \citep{Heller2016} is used to fit an ellipse to each individual cell contour and extract the angle and magnitude of the longest axis of the ellipse.  The angle and magnitude pair are then used to define the main axis of expansion for each cell in the modeling domain (Figure \ref{Exp_images}F).  For simulated cells, the expansion direction is calculated similarly to experimental images using resulting cell shapes from the final time step of each simulation (Figure \ref{Exp_images}G).
	
	Curvature of the L1 layer of the SAM in both experiments and simulations is quantified for comparison and model validation. For both experimental and simulation images, the center of each cell in the L1 layer is recorded and a circle is fit to the resulting set of data points using the Circle Fit (Pratt method) in matlab \citep{MATLAB} (See SI 1.1).  The radius of the fitted circle is used as an approximation for the radius of curvature of the L1 layer of the SAM.

\subsection{Model Assumptions and Calibration}
\label{modelCalib}

\subsubsection{Mechanical Properties of Individual Cells}

Model parameters representing cell wall mechanical properties were calibrated using biophysical measurements from a large body of literature \citep[for reviews see][]{Geitmann2006,RoutierKierzkowska2013}. Model parameters determining the spatial distribution of WUS in simulations were calibrated using experimental data obtained by our group.

\begin{figure}
\includegraphics[width = \textwidth]{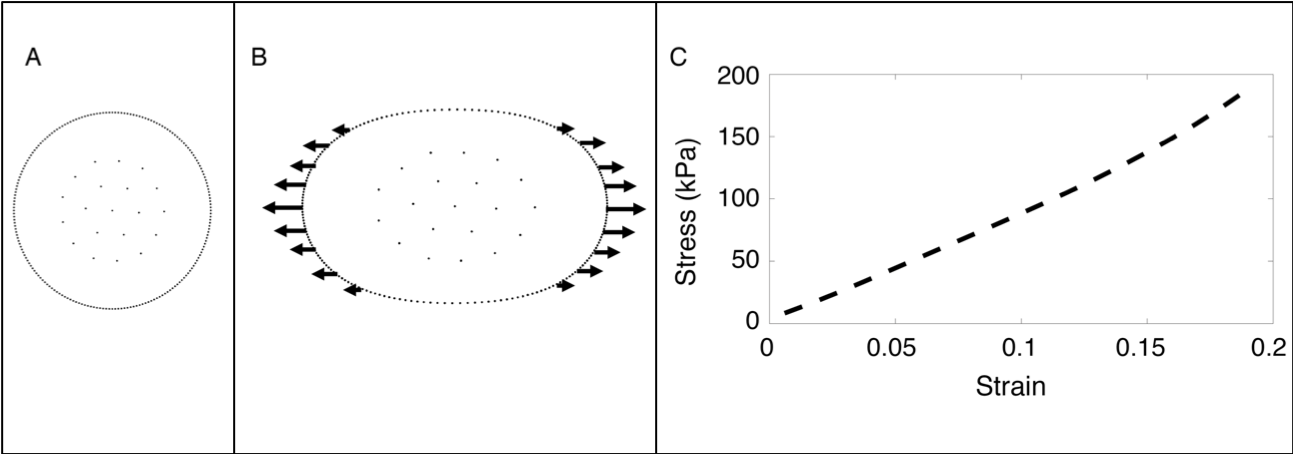}
\caption{Calibration of model parameters. (A-B) Calibration test to determine parameters for cell elasticity.  (A) Cell at equilibrium with no force applied.  (B) Cell has deformed after linearly increasing force is applied to nodes on both sides. (C) Stress versus strain graph for single cell calibration of modulus of elasticity.}
\label{Calibration_image}
\end{figure}

	Cell wall mechanical stiffness was calibrated using experimentally measured modulus of elasticity $(E)$ of a single cell.  Several different experiments have been performed to determine biological ranges for $E$ in plants \citep{RoutierKierzkowska2013}.  In model simulations, the modulus of elasticity is determined by applying a linearly increasing force to cell wall nodes on both sides of a cell and calculating the cell's deformation (Figure \ref{Calibration_image}A-B).  The slope of the graph of the stress versus strain curve provides the elasticity of the cell (Figure \ref{Calibration_image}C).  We have chosen values for $k_{bend}$ and $k_{linear}$ so that $E$ lies within the biological range of $(.1-1)$ MPa measured for plant cells \citep{Geitmann2006}.  
	
\begin{figure}
\includegraphics[width = \textwidth]{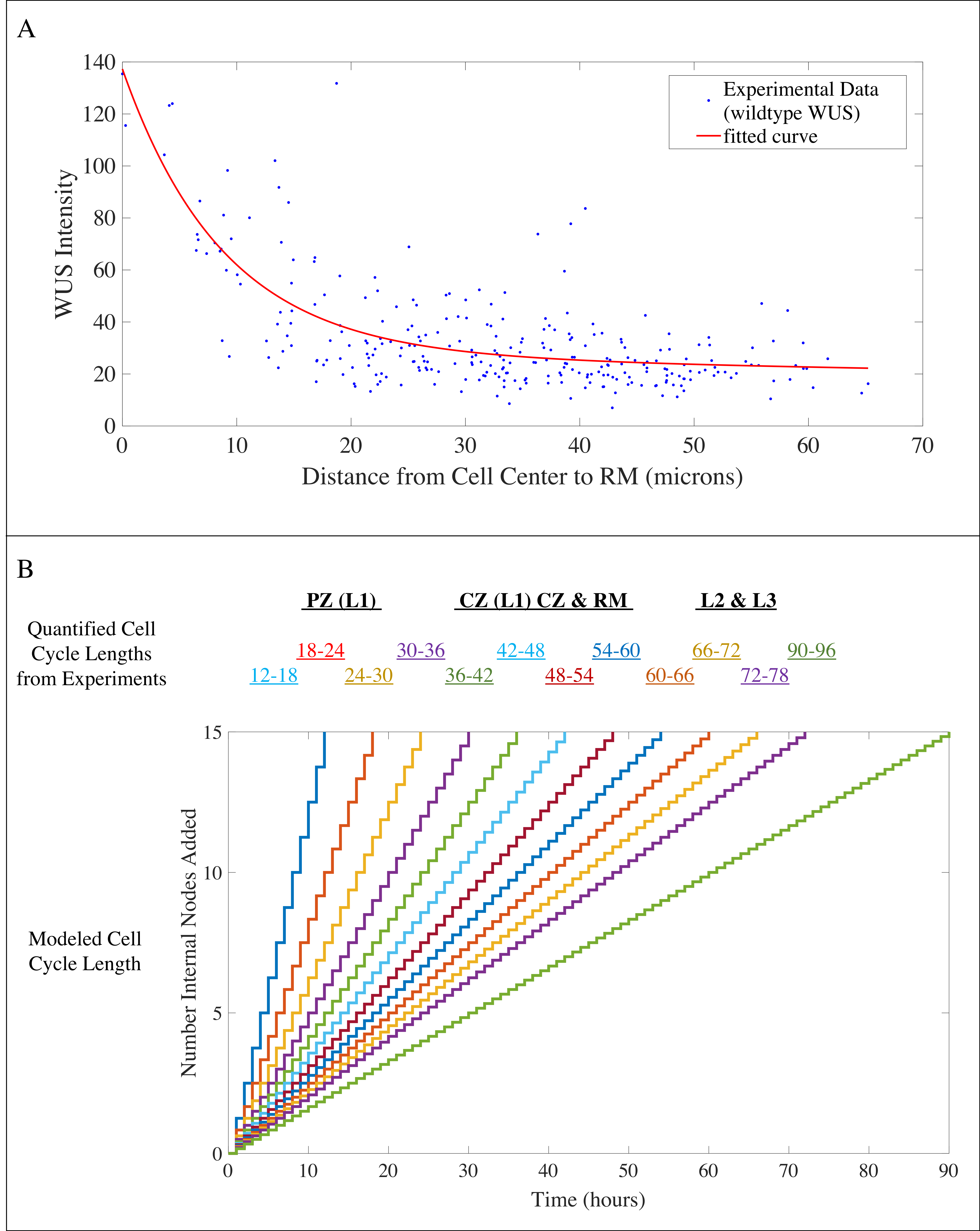}
\caption{(A) Graph showing the levels of WUS protein distribution in space. The WUS levels in different cells are plotted as a function of the distance from the RM. Blue dots represent experimentally quantified WUS levels.  Red line represents the fitted curve from equation \ref{WUS_dist}.  
(B) Graph showing the frequency of addition of internal nodes based on cell cycle length.  Cell growth rates are assumed to be directly correlated to the cell cycle length derived from experimental observations in an earlier study (Reddy et al., 2004).}
\label{Calibration_image_growth}
\end{figure}

\subsubsection{Cell Growth Rates}
	The spatial confinement of WUS to an exact domain within the SAM was shown to be crucial for maintaining a constant number of stem cells over time \citep{Yadav2010}.  In the model, the WUS density distribution is created by assigning each cell an average concentration of WUS determined as follows.  Experimental images of 13 different meristems were used to derive a function for the average WUS concentration of each individual cell based on the distance from its cell center to the RM where \textit{WUS} is expressed (Figure \ref{Calibration_image_growth}A). 
	
	This data was fit to an exponential function because WUS quantification from experiments suggests that the WUS protein distribution is exponentially distributed (Figure \ref{Calibration_image_growth}A).  \textit{WUS} is expressed in a few cells of the rib meristem (RM) called the niche/Organizing Center (OC) located just beneath the CZ and migrates from the RM into adjacent cells.  Since the distribution of the WUS signal from experiments is exponential, and WUS signaling dynamics have been previously modeled using reaction-diffusion equations \citep{Jonsson2005,Yadav2011}, we chose to fit an exponential function to the experimental data.  This resulted in the following concentration of WUS for an individual cell:

\begin{equation}
\label{WUS_dist}
WUS(x) = 109.6*\exp(-0.1135*x) + 27.69*\exp(-0.003414*x)
\end{equation}
where $x$ is the distance from the cell center to the RM where \textit{WUS} is expressed.

	The growth rate of each cell is determined in the model by its WUS concentration (See Table \ref{tab:3}).  Several experimental observations suggest that higher levels of WUS may inhibit cell growth and lower levels promote cell growth \citep{Reddy2004}.  These observations show that: a) WUS protein accumulates at higher levels in the slow growing RM and the CZ, and at lower levels in the fast growing PZ; b) ectopic activation of WUS outside the RM destabilizes WUS leading to a lower accumulation and increased growth rates; c) ectopic overexpression of a nuclear-enriched form of WUS leads to highly irregular SAMs which could be due to local differences in WUS concentration that in turn inhibit or stimulate growth in adjacent cells/regions \citep{Perales2016,Snipes2018,Yadav2010}.  Therefore, we assume in the model that  cells with the lowest concentration of WUS have the highest growth rate and cells with the highest concentration of WUS have the lowest growth rate (Figure \ref{Calibration_image_growth}B and Table \ref{tab:3}).  

\begin{table}
\renewcommand{\arraystretch}{1.25}
\caption{Cell cycle length as a function of WUS intensity.  Data taken with permission from Reddy et al. \citeyearpar{Reddy2004}.}
\label{tab:3}       
\begin{tabularx}{\textwidth}{| X | X |}
\hline
\textbf{WUS Intensity} & \textbf{Cell Cycle Length}\\
\hline
WUS $\leq$ 12 & 12-18 hours\\
\hline
12 $<$ WUS $\leq$  24 & 18-24 hours\\
\hline
24 $<$ WUS $\leq$ 36 & 24-30 hours\\
\hline
36 $<$ WUS $\leq$ 48 & 30-36 hours\\
\hline
48 $<$ WUS $\leq$ 60 & 36-42 hours\\
\hline
60 $<$ WUS $\leq$ 72 & 42-48 hours\\
\hline
72 $<$ WUS $\leq$ 84 & 48-54 hours\\
\hline
84 $<$ WUS $\leq$ 96 & 54-60 hours\\
\hline
96$<$ WUS $\leq$ 108& 60-66 hours\\
\hline
108$<$ WUS $\leq$ 120 & 66-72 hours\\
\hline
120$<$ WUS $\leq$ 132 & 72-78 hours\\
\hline
132 $<$ WUS & 90-96 hours\\
\hline
\end{tabularx}
\end{table}

\subsubsection{Boundary Conditions}
	
	There is only one boundary condition imposed during simulations.  The bottom most layer of cells in the deeper L3 layers has a higher damping coefficient and subsequently this layer of cells acts as a barrier the same way the cells of the stem would in the biological system (Table \ref{tab:2}).  Other cells in the tissue move and fluctuate freely as cells grow and interact.  WUS concentration of each cell is initially determined by Eq. \ref{WUS_dist} (Figure \ref{Calibration_image_growth}A and section \ref{modelCalib}) based on the location of the center of each cell.  Since we assume a steady-state distribution of WUS, once the WUS signaling domain is set up upon initiation of the tissue, the WUS concentration of each cell is not updated and therefore the WUS signaling domain will also move and fluctuate freely as cells grow and interact.

\subsubsection{Timescale}

	In an unperturbed system, spatial domains of chemical signaling remain unchanged and balanced by underlying feedback mechanisms \citep{ReddyandMeyer2005}.  Thus, in our model, we assume that steady-state, spatial distribution of WUS is maintained over the simulation time period and therefore we do not take into account transcription factor and protein movement explicitly.  In plants treated with ectopic activation of CK signaling, obvious changes in the size of the WUS signaling domain and shape of the meristem occurred by 22 hours after treatment \citep{Snipes2018}.  For this reason, the time period of 20 hours for simulations was chosen because it was long enough to observe the impact of signaling changes on cell growth rates as well as determine how these changes translate into changes in tissue morphology.
		
	Cell proliferation rates and division plane orientation affect both the shape and size of individual cells as well as topology of the tissue.  Coordinated division plane orientation and expansion of cells as a mechanism for determining shape of the tissue is especially important in plants since cells do not rearrange.  In addition, creation of new cell walls leads to local reinforcement of the tissue altering mechanical properties of the tissue in a preferential direction.  However, the position of the new cell wall is ultimately determined by the preprophase band (PPB), a ring structure formed by cortical microtubules before the cell enters mitosis. 

	 Previous experimental studies suggest that cortical microtubules orient according to the maximal mechanical stress direction which is largely determined by tissue shape \citep{Hamant2008,Sampathkumar2014a,Sampathkumar2014b, Uyttewaal2012,Williamson1990}.  This suggests that cortical microtubules serve as intermediates between tension patterns in the tissue  in cell walls and cell division plane orientation.   Thus, distribution of stress throughout the tissue provides supracellular cues that play a role in determining division plane orientation of individual cells.
	
	For this reason, simulations in this paper do not include division.  Simulations in this paper test mechanisms for SAM growth based on the combined contribution of mechanical properties of sub-cellular components of individual cells via anisotropic cell growth directions and varied cell growth rates based on WUS concentration and predict how these specific mechanisms establish the distribution of stress throughout the tissue.  In order to run simulations that quantify the relative contributions of chemical versus mechanical signaling in determining division plane orientation, it is necessary to first gain biological insights about individual cell mechanisms for anisotropic cell expansion as well as mechanical interactions between neighboring cells before new cells are added to the tissue. 

	 Future simulations will encompass a larger timescale and include cell division to predict new mechanisms for SAM development that quantify  relative contributions of chemical versus mechanical signaling in determining division plane orientation.  The extended model will provide a platform for testing the feedback between mechanical properties of the tissue that contribute to cell division orientation patterns and cell division orientation patterns that affect mechanical properties of the tissue.

\section{Results}
\label{results}
\subsection{Factors Determining Overall Shape of the SAM}
\label{baseline}
\begin{figure}
\centering
\includegraphics[width = \textwidth]{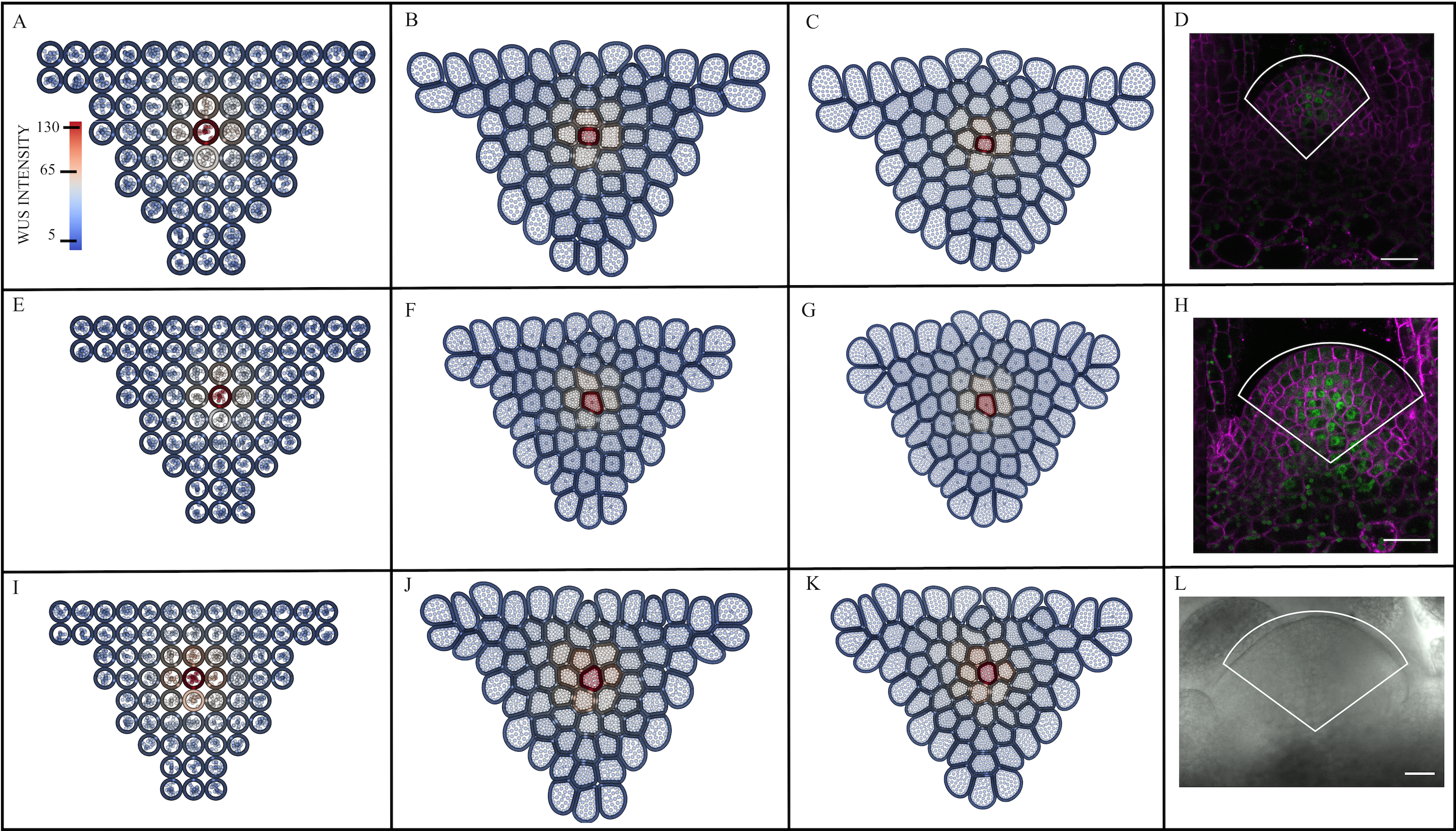}
\caption{Time snapshots of simulations of the formation of the shape and structure of the SAM of  \textit{Arabidopsis} and experimental images. (A-C) Simulation of wildtype SAM growth with diameter of CZ equal to $15 \mu$m and resulting radius of curvature of the L1 layer equal to $51.27 \mu$m. (D) Experimental image of wildtype SAM obtained by our group. (E-G) Simulation of SAM growth with diameter of CZ equal to $34 \mu$m and radius of curvature equal to $39.38 \mu$m.  (H) Experimental image of meristem experiencing the ectopic overactivation of CK signaling in the CZ for 12 hours obtained by our group. (I-K) Simulation of SAM growth with diameter of CZ equal to $56 \mu$m and radius of curvature equal to $86.42 \mu$m. (L) Experimental image of meristem tagged with a potent nuclear localization signal (nls-eGFP-WUS).  In (D),(H) and (L) the simulation domain is shown in the enclosed ares in white.  Scale bars are $20\mu$m}
\label{Time_series}  
\end{figure}

The computational model was used to study morphological implications of individual cell behaviors in the SAM by simulating combined growth of the L1, L2, and deeper L3 layers.  Model simulations were run to determine whether layer dependent mechanical anisotropy at the sub-cellular and cellular level combined with experimentally calibrated diameter of the WUS signaling domain were sufficient to reproduce experimentally observed expansion directions of cells as well as experimentally observed shape and size of the SAM characterized by curvature of the L1 layer (Figure \ref{Time_series}).  In addition, model predictive simulations were run to test the hypothesis that WUS concentration of individual cells controls individual cell growth rates as a mechanism for generating SAM shape and structure.

	For simulations of wildtype SAM growth (Figure \ref{Time_series}A-C), the following assumptions were made. The diameter of the CZ and resulting WUS signaling domain were calibrated using experimental data and described by Eq. \ref{WUS_dist} (Figure \ref{Calibration_image_growth}A and section \ref{modelCalib}).  Cell growth rates were determined based on the WUS concentration of individual cells (See Table \ref{tab:3}, Figure \ref{Calibration_image_growth}B and section \ref{modelCalib}). Lastly,  cells in the L1 and L2 layers were assigned growth direction vectors parallel to the surface of the SAM and all cells in the deeper layers were assigned growth direction vectors perpendicular to the surface of the SAM.

\begin{figure*}
\centering
\includegraphics[width = .85\textwidth]{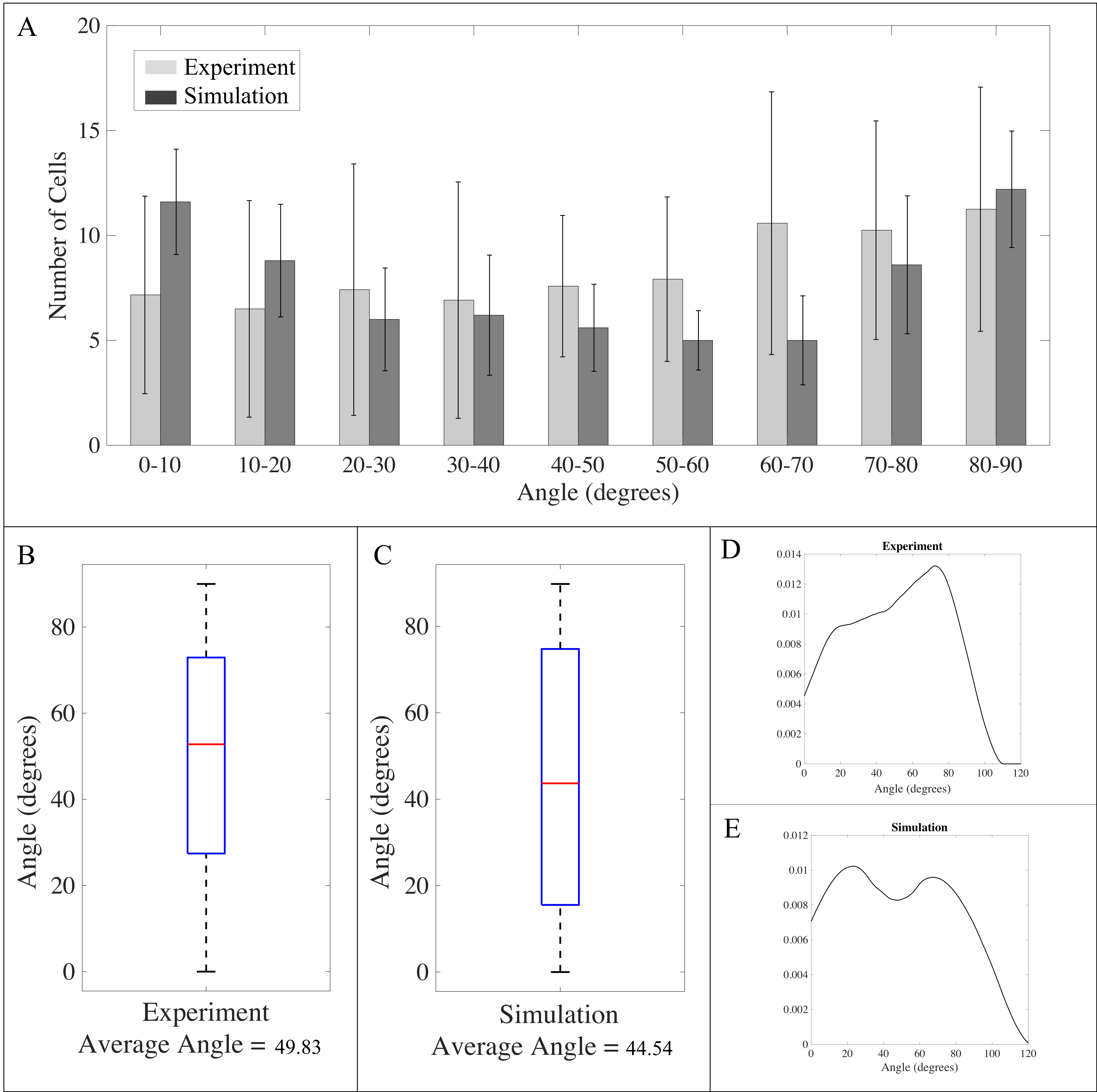}
\caption{Model validation for simulating wildtype tissue growth. (A) Distribution of the angle of the main axis of expansion of cells in experiments versus simulations.  Boxplots showing average angle of the main axis of expansion of cells in wildtype experiments (B) and simulations (C).  Kernel density estimation (KDE) plots for the angle of the main axis of expansion of cells across experiments (D) and computational simulations (E) respectively.  KDE plots demonstrate both data sets follow a bimodal distribution with one mode close to 90 degrees and the other mode close to 0 degrees.}
\label{wildtype_angles}  
\end{figure*}

	Tissue shapes obtained in wildtype simulations were compared with tissue shapes observed in experimental images. Namely, the distribution of the angles of the main axis of expansion for all cells in the tissue (See section \ref{expMethods}) obtained in simulations and experiments were compared to quantify the impact of the expansion direction of individual cells on overall tissue shape (Figure \ref{wildtype_angles}A-E).  A kernel density estimation (KDE) plot for the angle of the main axis of expansion of cells across 13 experimental images was compared to a KDE plot for the angle of the main axis of expansion of cells across 5 simulations.  KDE plots demonstrate that both data sets follow a bimodal distribution with one mode close to 90 degrees and the other mode close to 0 degrees.  These results are consistent with experimental observations wherein cells in the deeper L3 layers expand perpendicular to the surface of the SAM, i.e. the main axis of expansion is 90 degrees, and cells in the L1 and L2 layers expand parallel to the surface of the SAM, i.e. the main axis of expansion is 0 degrees.  Comparison between KDE plots for experimental and simulation data indicate that there was not a significant difference between the two groups.  Thus, model assumptions used in the wildtype simulations were enough to reproduce the average angle for the main axis of expansion seen in experimental images. 
	
	Lastly, we demonstrated that model simulations reproduced experimentally observed curvature of the L1 layer of the SAM (Figure \ref{curvatures}).  The average radius of curvature of the L1 layer of the SAM was computed from single-time-point experimental images of 13 different wildtype plants as well as data output from the last time step of 5 wildtype simulations. A t-test comparing the average radius of curvature from wildtype experimental images (50.75 $\mu$m) to the average radius of curvature from wildtype simulations ($67.19 \mu$m) resulted in $p = 0.0656$ demonstrating that there was no significant difference between simulations and experimental data ($\alpha = 0.05$).	

\subsection{Impact of WUS Concentration of Individual Cells Controlling Cell Growth Rates on Overall Shape of SAM}
\label{cellGrowth}

\begin{figure*}
\centering
\includegraphics[width =\textwidth]{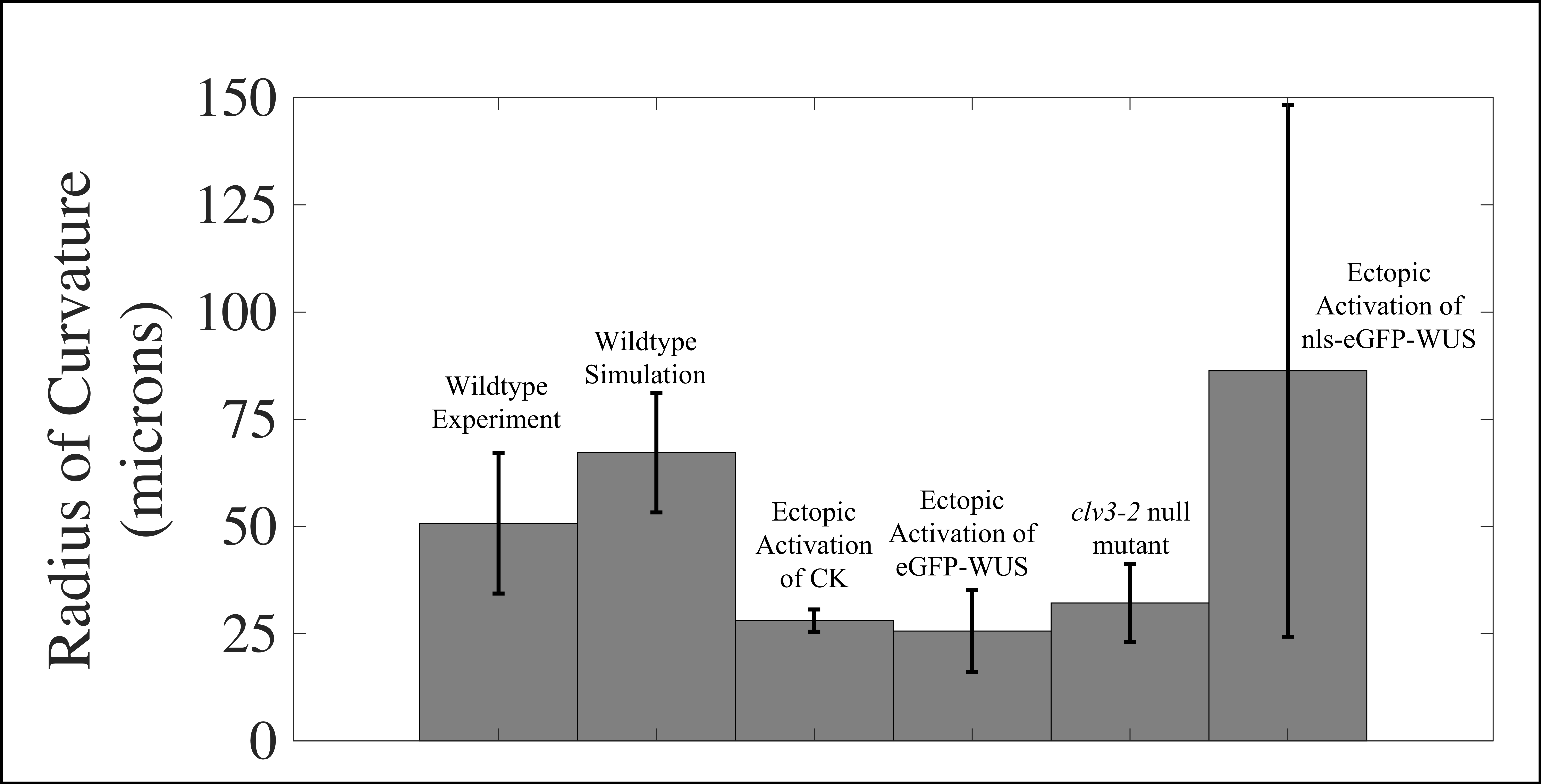}
\caption{Comparison of radii of curvature for wildtype meristems from experiments and simulations and radii of curvature for four alternative systems.  Average radius of curvature across 13 wildtype experimental meristems is 50.75 $\mu$m, average radius of curvature across 5 wildtype simulations is 67.19 $\mu$m, average radius of curvature across 26 ectopic activation of CK experimental meristems is 28.06 $\mu$m, average radius of curvature across 7 ectopic activation of eGFP-WUS experimental meristems is 25.63 $\mu$m, average radius of curvature across 8 \textit{clv3-2} null mutant experimental meristems is 32.17 $\mu$m and average radius of curvature across 10 ectopic activation of nls-eGFP-WUS experimental meristems is 86.28$\mu$m.}
\label{curvatures}  
\end{figure*}

In addition to quantifying wildtype SAM growth described above, the average curvature of the L1 layer of the SAM was computed from experimental images from 26 ectopic activation of CK experimental meristems (avg = 28.06 $\mu$m), 7 ectopic activation of eGFP-WUS experimental meristems (avg = 25.63 $\mu$m), 8 \textit{clv3-2} null mutant experimental meristems (avg = 32.17 $\mu$m) and 10 ectopic activation of nls-eGFP-WUS experimental meristems (avg = 86.28$\mu$m) (Figure \ref{curvatures}).  A t-test comparing the average curvature of the L1 layer of wildtype meristems to each of the four alternative systems resulted in $p = 3.0230\textrm{e}-08$,  (ectopic activation of CK), $p = 0.0016$ (ectopic activation of eGFP-WUS), $p= 0.0616$ (ectopic activation of nls-eGFP-WUS) and $p = 0.0060$ (\textit{clv3-2} null mutants), respectively.  These results demonstrate that ectopic activation of CK meristems, ectopic activation of eGFP-WUS meristems and \textit{clv3-2} null mutants all lead to significant increase in the curvature of the L1 layer of the SAM and ectopic activation of nls-eGFP-WUS meristems are not significantly more curved than wildtype meristems ($\alpha = 0.05$).

\begin{figure*}
\includegraphics[width = \textwidth]{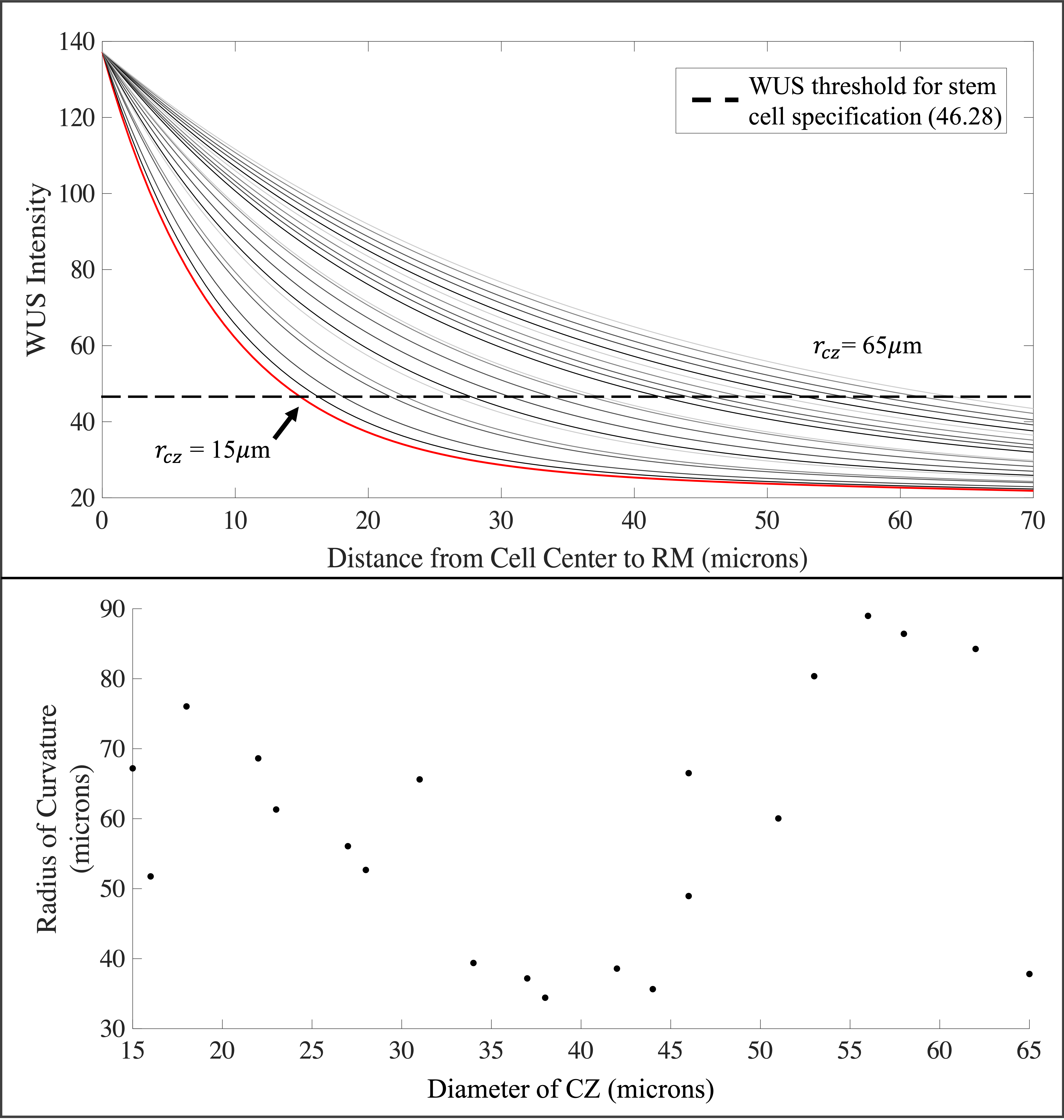}
\caption{Impact of WUS specified stem cell identity on overall tissue shape of SAM.  (A) The twenty different functions used as input for the WUS signaling domain in simulations where diameter of CZ is varied.  Red line is function used as input for WUS signaling domain in wildtype simulations.  Dashed line is WUS threshold for stem cell specification, i.e. cells whose WUS concentration falls above the red line behave as stem cells in simulations. (B) Resulting curvature of the L1 layer of the SAM for each choice of diameter of the CZ from the twenty different simulations. First data point with diameter of the CZ equal to $15\mu$m is average curvature of the L1 layer over five wildtype simulations.}
\label{variation}  
\end{figure*}

	To investigate the impact of WUS concentration of individual cells controlling cell growth rates on curvature of the L1 layer, twenty simulations were run with different diameters of the CZ (Figure \ref{variation}). Values for the diameter of the CZ were chosen from the range $15 \mu$m - $65\mu$m.  This range was used because the average diameter of the CZ in wildtype experimental images is 15$\mu$m, and the maximum possible diameter of the CZ for simulations is 65$\mu$m.  For sampling, the range $15 \mu$m - $65\mu$m was divided into twenty intervals and each interval was sampled exactly once (without replacement), so that the entire range for the parameter was explored. Each of the twenty samples was used to generate a different WUS signaling domain for a new simulation (Figure \ref{variation}). 
	
	Results demonstrate that the relationship between the diameter of the CZ and radius of curvature of the L1 layer of the SAM is not linear.  Meristems with diameter of the CZ between $32\mu$m and $45\mu$m have the smallest radius of curvature.  In addition, once the diameter of the CZ passes $45\mu$m, meristem growth starts to flatten out and the radius of curvature of the L1 layer increases.  Model predictive simulations demonstrating significant morphological changes due to WUS concentration of individual cells controlling growth rates could be linked to WUS concentration-dependent transcriptional regulation of \textit{CLV3} \citep{Perales2016} (See section \ref{diss} for details). Results from each of the twenty different simulations along with the WUS signaling domain used in each of the twenty different simulations are provided in Figure \ref{variation}.
	
	Individual cell growth rates were assigned as before (See section \ref{modelCalib} and Table \ref{tab:3}) and layer dependent mechanical properties of cells remain the same.

\subsection{Impact of WUS Concentration of Individual Cells Controlling Cell Growth Rates on Internal Pressure Distribution in Tissue}
\label{stress}
\begin{figure*}
\includegraphics[width = \textwidth]{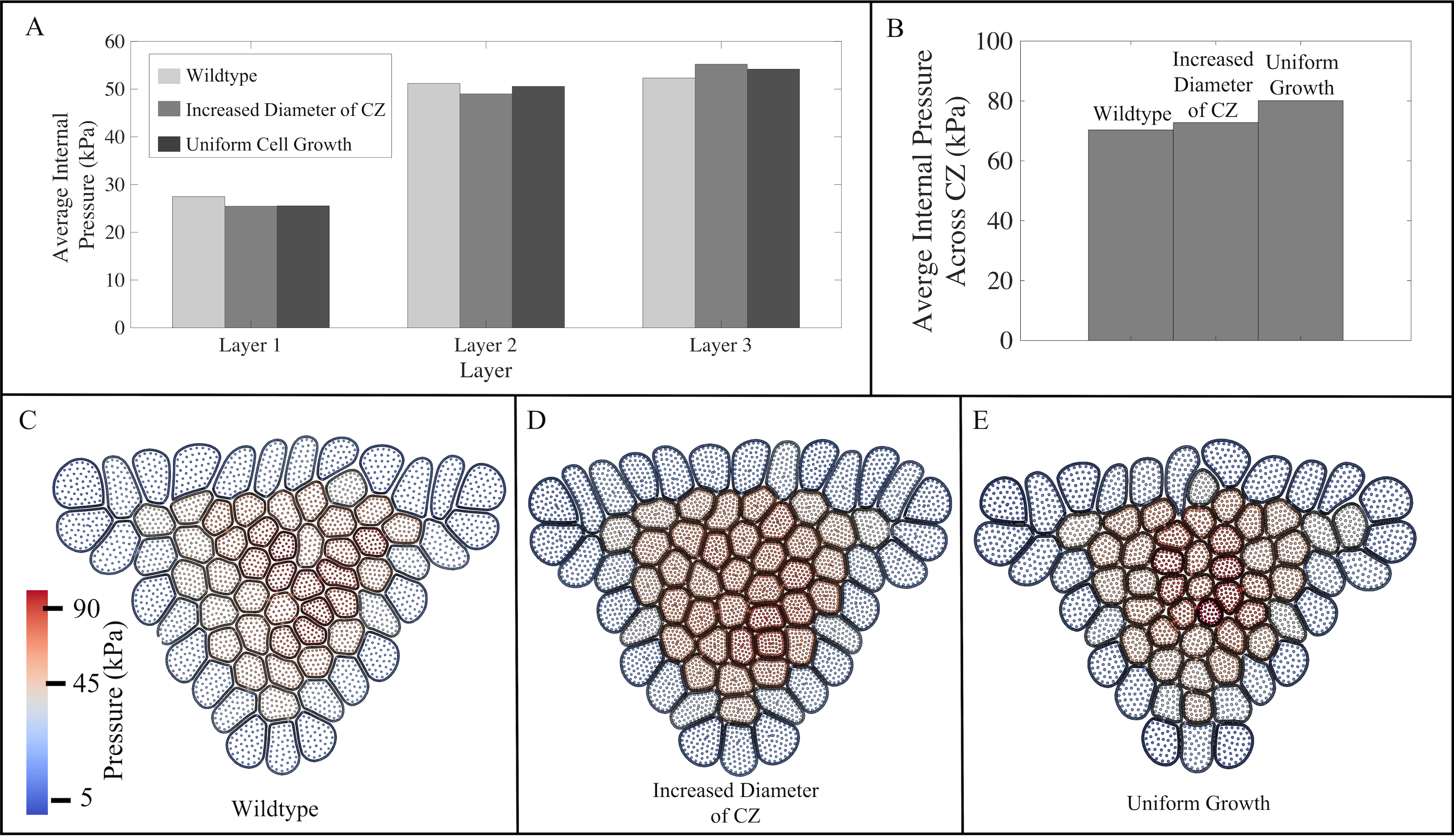}
\caption{Change in pattern of distribution of internal pressure across three different simulations.  (A) Distribution of internal pressure across L1, L2 and L3 layers from wildtype, increased diameter of CZ (34 $\mu$m $\leq$ diameter $\leq$ 44 $\mu$m), and uniform cell growth (diameter of CZ equal to 56$\mu$m) simulations.  (B) Distribution of internal pressure across CZ from wildtype (avg = 70.32 kPa), increased diameter of CZ (avg = 72.77 kPa), and uniform growth simulations (avg = 80.09 kPa). (C) Distribution of internal pressure in wildtype simulation.  (D) Distribution of internal pressure in increased diameter of CZ simulation. (E) Distribution of internal pressure in uniform growth simulation.}
\label{Pressure_plots}  
\end{figure*}

 The average internal pressure of individual cells across the L1, L2 and deeper L3 layers of the SAM was calculated after 20 hours of growth separately in simulations representing wildtype (diameter of CZ equal to $15 \mu$m), increased diameter of CZ (34 $\mu$m $\leq$ diameter $\leq$ 44 $\mu$m), and uniform cell growth (diameter of CZ equal to 56$\mu$m) (Figure \ref{Pressure_plots}A) simulations.  Next, the average internal pressure across the CZ was calculated for each simulation (Figure \ref{Pressure_plots}B). Results show distinct patterns of pressure accumulation for wildtype (avg = 70.32 kPa), increased diameter of CZ (avg = 72.77 kPa) and uniform cell growth simulations (avg = 80.09 kPa)  (Figure \ref{Pressure_plots}C-E).  Stem cells in uniform growth simulations experience higher pressure compared to wildtype and increased diameter of CZ simulations.  Model predictive simulation results suggest that distribution of pressure in the tissue could play a role in controlling the rate of cell growth (See section \ref{diss} for details).
	
\section{Discussion and Conclusions}
\label{diss}

The growth and development of the SAM depend on spatial and temporal coordination of cell growth patterns, anisotropic cell wall mechanical properties, as well as chemical and mechanical signaling feedbacks controlling cell behavior. In this paper, a novel cell-based, SCE model is presented and used for studying morphological implications of individual cell behaviors by analyzing the combined impact of WUS concentration of individual cells controlling cell growth rates and mechanical properties of sub-cellular components of individual cells and the cell wall on the shape of the SAM characterized by the curvature of the L1 layer.  

	The main novelty of this paper is the extension of the general SCE approach to develop a detailed, biologically-calibrated model describing the dynamics of the three layers of the SAM that tests impact of the combined chemical and mechanical effects on regulating SAM growth and shape.  The model combines detailed representations of cell wall mechanical properties controlling anisotropic cell expansion, deformation of the middle lamella of the cell wall, and increase in cytoplasmic pressure to generate turgor pressure, as well as dynamic interactions between these different sub-cellular components.

	In section \ref{baseline}, model simulations were shown to successfully reproduce emergent properties of the multi-layered SAM tissue including the main axis of expansion of the tissue and average curvature of L1 surface layer of the SAM that matched experiments (Figure \ref{wildtype_angles}).  This provides evidence in support of the hypothesized mechanism of SAM shape formation based on combining layer dependent mechanical anisotropic distribution at the sub-cellular and cellular level with experimentally calibrated diameter of the CZ determining individual cell growth rates as a function of WUS concentration.

	In section \ref{cellGrowth}, the model was used to successfully test the new hypothesis that WUS concentration of individual cells could impact SAM shape.  One of the novel features of the model is the separate representation of individual cells, including cells in the L3 and deeper layers.  This makes it possible to test hypotheses about the role of WUS concentration in impacting cell behaviors directly or indirectly by specifying cell identity, especially in the deeper layers where it is difficult to experimentally track cells over time.   
	
	Model predictive simulations demonstrate that significant morphological changes during SAM growth were associated with changes in the diameter of the CZ (Figure \ref{variation}).  Moreover, the simulations of WUS concentration-dependent growth could also be linked to its concentration-dependent transcriptional regulation of \textit{CLV3} \citep{Perales2016}.  This is because earlier analysis revealed that WUS activates \textit{CLV3} transcription at lower concentrations and represses \textit{CLV3} transcription at higher concentrations \citep{Perales2016}.  In addition, \textit{CLV3}-mediated signaling is required for nuclear accumulation of WUS in the CZ.  Perhaps, \textit{CLV3}-mediated signaling enriching WUS in the nuclei of CZ cells could restrict growth, while the cells in the PZ that are displaced out of the \textit{CLV3}-signaling zone accumulate lower nuclear WUS and divide faster. 
	
	Predictive simulations reveal that meristems with a CZ diameter between $32\mu$m-$45\mu$m have a smaller radius of curvature than meristems with a higher CZ diameter ($>45\mu$m) (Figure \ref{variation}). These results are consistent with experimental observations wherein ectopic activation of eGFP-WUS in the CZ led to overall lower WUS and an increase in CZ diameter along with an enlarged and pointy meristem (Figure \ref{Exp_images}D and \ref{variation}). Whereas, patches of higher WUS accumulation observed in meristems experiencing ectopic activation of nls-eGFP-WUS led to flatter and irregularly shaped SAMs which could be due to heterogeneity in growth rates and may also be due to the loss of CZ identity in patches (Figure \ref{Exp_images}E and \ref{variation}) \citeyearpar[For details see Figure 5C and F in Perales et al.][]{Perales2016}.  Though it is unclear whether high WUS concentration of individual cells restricts growth directly or indirectly by specifying cell identity,  simulation assumptions that WUS concentration of individual cells controls growth rates are supported by an earlier study from our group showing that direct misexpression of WUS in the PZ leads to retardation of growth \citep{Yadav2010}.  Thus, future time-series data from experiments in combination with additional modeling studies are required to uncouple the impact of cell identity and WUS concentration on growth rates.
	
	In section \ref{stress}, predictive model simulations revealed that changes in the size of the diameter of the CZ resulted in distinct distributions of internal cell pressure across the stem cell niche (Figure \ref{Pressure_plots}).  Namely, increasing the diameter of the CZ from $15 \mu$m observed in wildtype experiments to $65 \mu$m in uniform growth simulations where every cell in the tissue behaves like a stem cell, increased pressure in the CZ by 10 kPa.  Model simulation results indicate that cell behavior in response to changes in internal cell pressure could provide an additional mechanism for maintaining the correct ratio of slow growing cells in the CZ to fast growing cells in the PZ resulting in a stable population of stem cells and the correct shape and size of the meristem.  More specifically, distribution of pressure in the tissue could play a role in controlling the rate of cell growth and division, i.e. stem cells under higher pressure in the CZ may divide less frequently than differentiated cells under lower pressure in the PZ \citep{Vollmer2017}.
		
	To summarize, we demonstrated using a cell-based model how layer dependent anisotropic mechanical properties of sub-cellular components of individual cells and the cell wall and WUS concentration of individual cells control cell behavior and ultimately determine the final size and shape of the meristem. Many persisting questions about interactions between chemical and mechanical signaling can be studied using further extensions of the model.  
	
	In particular, we plan to extend the model by combining the mechanical sub-model with a dynamic signaling model.  Understanding how cell growth rates, cell size, cell shape and cell division patterns facilitate signaling diffusion is crucial for gaining a better understanding of the spatio-temporal regulation of the stem cell niche.  For example, the extended model can be used to test the hypothesis that division plane orientation impacts diffusion by the creation of new plasmodesmata in a preferential direction when new cell walls are laid down.  If the majority of cells in the deeper L3 layers divide periclinally, the creation of new plasmodesmata along the apical-basal axis of the meristem   would create a vertical path for diffusion.  
	
	Alternatively, there is evidence that cell wall stiffening may prevent diffusion through plasmodesmata \citep{Daum2014}.  Performing \textit{in silico} experiments that test the role of plasmodemata-mediated regulation of WUS diffusion in controlling WUS levels could lead to new insights into the plasmodesmata distribution and conductance properties which are otherwise challenging to determine experimentally.
	
	In addition, combining the mechanical sub-model with a dynamic signaling model would make it possible to link sub-cellular processes regulating intracellular WUS distribution to its spatial accumulation and the regulation of \textit{CLV3} transcription. It will also enable us in the future to test the relative roles of WUS, CK and mechanical signals in determining the growth rates and division plane orientation of individual cells. Moreover, new insights as to how cells within a tissue determine the orientation of their plane of division would make it possible to study the effect of division plane orientation on morphological features such as cell growth direction and curvature of the L1 layer of the SAM.
	
\section{Data Accessibility}
\label{code}
A flowchart outlining the computational implementation of the model is shown in Figure S1.1.  In addition, code and several simulation movies can be found at:\\
https://sites.google.com/view/mikahlbanwarthkuhn/research/plant-stem-cells.

\begin{acknowledgements}
The authors acknowledge partial support from the National Science Foundation Grant DMS-1762063 through the  joint NSF DMS/NIH NIGMS Initiative to Support Research at the Interface of the Biological and Mathematical Sciences. G.V.R. was also partially supported by the  National Science Foundation Grant IOS-1456725 and RSAP-AES mission funding to GVR.  CGR acknowledges funding from National Science Foundation Grant MCB-1716972.  The authors thank Andrew Whitaker for his invaluable help with transferring computational code to C++ language and code optimization.  The authors thank Dr. Weitao Chen for her invaluable help with many computational problems as well as insightful discussions about model development. 

\end{acknowledgements}
\bibliographystyle{spphys}  
 \bibliography{references}

\end{document}